\documentclass[12pt,letterpaper]{JHEP3}
\usepackage{amsmath,amssymb,amsfonts,xspace}
\usepackage{epsfig}

\numberwithin{equation}{section}

\def\varpi{{\bf z}}

\def\sign{{\rm sign}}

\def\Im{\,{\rm Im}\,}
\def\Re{\,{\rm Re}\,}

\def\({\left(}
\def\){\right)}
\def\[{\left[}
\def\]{\right]}
\def\hf{{1\over 2}}

\newcommand{\tX}{\tilde{X}}

\renewcommand{\d}{\mathrm{d}}
\newcommand{\de}{\mathrm{d}}

\newcommand{\I}{\mathrm{i}}
\newcommand{\e}{\mathrm{e}}
\newcommand{\cL}{\mathcal{L}}

\newcommand{\p}{\partial}

\newcommand{\cD}{\mathcal{D}}
\newcommand{\cF}{\mathcal{F}}

\newcommand{\cG}{\mathcal{G}}
\newcommand{\cK}{\mathcal{K}}
\newcommand{\cM}{\mathcal{M}}

\newcommand{\cE}{\mathcal{E}}

\newcommand{\CX}{\mathcal{X}}
\newcommand{\cR}{\mathcal{R}}
\newcommand{\cT}{\mathcal{T}}

\newcommand{\zb}{\bar{z}}

\newcommand{\rmd}{\rm d}

\DeclareSymbolFont{AMSa}{U}{msa}{m}{n}
\DeclareSymbolFont{AMSb}{U}{msb}{m}{n}
\DeclareMathSymbol{\fieldR}{\mathalpha}{AMSb}{"52}

\newcommand{\N}{{\mathcal N}}
\newcommand{\kahler}{{K\"ahler}\xspace}

\newcommand{\cZ}{\mathcal{Z}}
\newcommand{\cI}{\mathcal{I}}
\newcommand{\cO}{\mathcal{O}}

\newcommand{\cU}{\mathcal{U}}
\newcommand{\cA}{\mathcal{A}}

\newcommand{\pa}{\partial}
\newcommand{\nn}{\nonumber}

\newcommand{\eps}{\epsilon}
\newcommand{\ve}{\varepsilon}
\newcommand{\IR}{\mathbb{R}}
\newcommand{\IC}{\mathbb{C}}
\newcommand{\IZ}{\mathbb{Z}}

\newcommand{\tzeta}{\tilde\zeta}

\newcommand{\txi}{\tilde\xi}

\newcommand{\CP}{\IC P^1}

\def\bea{\begin{eqnarray}}
\def\eea{\end{eqnarray}}
\def\be{\begin{equation}}
\def\ee{\end{equation}}
\def\ba{\begin{align}}
\def\ea{\end{align}}
\def\bse{\begin{subequations}}
\def\ese{\end{subequations}}

\def\bi{\bar \imath}
\def\bj{\bar \jmath}
\def\ba{\bar a}

\def\bz{\bar z}
\def\bY{\bar Y}

\def\bG{ \bar G }

\def\bF{\bar F}

\def\ui#1{^{[#1]}}
\def\di#1{_{[#1]}}

\def\txii#1{{\tilde\xi}^{[#1]}}
\def\ai#1{{\alpha}^{[#1]}}
\def\alpi#1{\alpha^{[#1]}}

\def\xii#1{\xi_{[#1]}}

\def\Ti#1{T^{[#1]}}

\def\Hij#1{H^{[#1]}}

\newcommand{\Li}{{\rm Li}}

\def\bpm{\IR_\mp}
\def\bmp{\IR_\pm}
\def\bp{\IR_-}
\def\bm{\IR_+}

\def\Kkl{\cK_{\gamma}}

\def\Ikl{\cI_{\gamma}}

\def\epskl{\epsilon_{\gamma}}
\def\Thhg{\Theta}

\def\Gg{\cG_{\gamma}}

\def\tve{\tilde{\varepsilon}}

\def\XXint#1#2#3{{\setbox0=\hbox{$#1{#2#3}{\int}$}
\vcenter{\hbox{$#2#3$}}\kern-.5\wd0}}

\def\hHij#1{H^{[#1]}}
\newcommand{\hU}{\hat{\mathcal{U}}}
\newcommand{\hCX}{\mathcal{X}}
\def\cij#1{c}
\def\ci#1{c}
\def\hn{n}
\def\hnkl{n_{\gamma}}

\def\ob{0_B}
\def\oa{0_A}
\def\opa{0_A^{\smash{_{'}}}}
\def\opb{0_B^{\smash{_{'}}}}

\newcommand{\cY}{\mathcal{Y}}

\def\hng#1{n_{\gamma_{#1}}}

\def\aA{\alpha_{\rm A}}
\def\aB{\alpha_{\rm B}}

\def\cLp{\cL_{\gamma}}
\def\cLm{\overline{\cL_{\gamma}}}

\title{
Quantum mirror symmetry and twistors
}
\preprint{LPTA/09-035, IPhT-T09/066}
\author{
Sergei Alexandrov$^1$,
Frank Saueressig$^2$
\\
$^1$ {\it Laboratoire de Physique Th\'eorique \&
Astroparticules, CNRS UMR 5207, \\
Universit\'e Montpellier II, 34095 Montpellier Cedex 05, France}\\
$^2$ {\it Institut de Physique Th\'eorique,
 CEA, IPhT, F-91191 Gif-sur-Yvette, France\\
CNRS, URA 2306, F-91191 Gif-sur-Yvette, France}\\
}
\abstract{Using the twistor approach to hypermultiplet moduli spaces,
 we derive the worldsheet, D($-1$),  and D1-instanton contributions to
 the generalized mirror map, relating Type IIA  and Type IIB string theory
 compactified on generic mirror Calabi-Yau threefolds.  For this purpose,
 we provide a novel description of the twistor space underlying the 
 Type IIB hypermultiplet moduli space where the SL(2, $\IZ$)-action
 is found to be free from  quantum corrections. The extent to which 
instanton effects may resolve  the perturbative singularities of the
 moduli space metric is discussed.
}

\begin{document}

\section{Introduction}
\label{sec_intr}

Instanton effects are one of the key ingredients in
understanding the mathematical structures of string theory
at the non-perturbative level. They dominate the dynamics
 at strong coupling and play a prominent role
in the non-perturbative dualities which are vital for
exploring in this regime.
Moreover, there is evidence that they 
are capable of resolving the singularities appearing in the
perturbative analysis of the string theory moduli space \cite{Ooguri:1996me}.

An important laboratory,
where such effects can be studied in detail,
is the compactification of Type II strings on a Calabi-Yau threefold (CY).
In this case the low-energy effective action (LEEA) has $N=2$ supersymmetry
and receives space-time instanton corrections
from Euclidean D-branes and NS5-branes wrapping supersymmetric
cycles of the internal space \cite{Becker:1995kb}.
Supersymmetry then dictates that the LEEA
is characterized by two manifolds: a special K\"ahler (SK) space $\cK$ determining
its vector multiplet sector, and a quaternion-K\"ahler (QK) space $\cM$ underlying its hypermultiplet
sector. For Type IIA strings compactified on a CY $X$ and Type IIB strings on $Y$, $\cK$ is given by
the K\"ahler moduli space $\cK_K(X)$ and the moduli space of complex
structures $\cK_C(Y)$, respectively:
\begin{center}
\begin{tabular}{|c|c|c|}
\hline
 & Type IIA$/X$ & Type IIB$/Y$ \\ \hline
$ \quad 
{\mbox{moduli}\atop \mbox{space}}\rule{0pt}{15pt}
\quad $ & $\qquad \cK_K(X) \times \cM_C(X) \qquad $
& $\qquad \cK_C(Y) \times \cM_K(Y) \qquad $
\\
\hline
\end{tabular}
\end{center}
At string tree-level, $\cM_C(X)$ and
$\cM_K(Y)$ can be obtained from $\cK_C(X)$ and $\cK_K(Y)$ via the
c-map \cite{Cecotti:1988qn,Ferrara:1989ik}. It thereby turns out that
the four-dimensional dilaton $\e^{\phi/2} \propto g_{(10)}$,
proportional to the ten-dimensional string coupling constant,
enters into $\cM$. Thus both hypermultiplet sectors
are subject to perturbative and non-perturbative string corrections.

A striking feature of these moduli spaces is
their conjectured relation via mirror symmetry \cite{Hori:2003ic}.
In its ``generalized'' or ``non-perturbative'' formulation, this duality
states that the moduli spaces arising from the compactification
of Type IIA strings on $X$ and Type IIB strings on $\tilde{X}$ (mirror to $X$) are actually identical.
On the vector multiplet side this amounts to the well-supported
classical mirror symmetry \cite{Candelas:1990rm},
which equates the complex structure moduli space of $X$ with
the complexified K\"ahler moduli space of $\tilde X$,
$\cK_C(X) = \cK_K(\tilde X)$. Generalized mirror symmetry
further implies the identification
\be\label{rel:mirror}
\cM_C(X) = \cM_K(\tilde X) \, ,
\ee
which is supposed to hold upon including the non-perturbative $g_s$-corrections
originating from D- and NS5-brane instantons \cite{Becker:1995kb,Ferrara:1995yx}.
In mathematical terms \eqref{rel:mirror} translates into an equivalence between elements of
the derived category $\cD(\tilde{X})$ of coherent sheaves (related to $\cM_K(\tilde X)$) and elements
in the derived Fukaya category $\cF(X)$ of SLAG submanifolds
(related to $\cM_C( X)$) (see \cite{Aspinwall:2004jr} for an introduction). In particular, it
encompasses the homological mirror symmetry conjecture \cite{MR1403918}.

At the level of the LEEA, mirror symmetry is realized via the
generalized (or non-perturbative) mirror map which relates the
physical fields of the Type IIA compactification, parameterizing $\cM_C(X)$, to the
physical fields of the Type IIB compactification, providing coordinates on $\cM_K(\tilde X)$.
The classical limit of this map has been obtained in \cite{Bohm:1999uk}.
Since the moduli spaces receive quantum corrections
from both worldsheet and D-brane instantons, it
is expected, however, that the full generalized mirror map
will also be subject to such corrections. The main result
of this paper, which can be found in eq.\ \eqref{mirtzetaz}, is the explicit construction
of this map, including the quantum corrections from worldsheet, D($-1$), and D1-instantons.

The derivation of the generalized mirror map requires the detailed knowledge
of the perturbative and non-perturbative corrections to $\cM$.
For the compactifications at hand these are given by perturbative worldsheet and $g_s$-corrections,
 contributions from D2-branes wrapping the SLAG three-cycles in Type IIA mirror to
 odd branes (D($-1$), D1, D3, D5) wrapping complex cycles of $\tilde{X}$ in Type IIB \cite{Ooguri:1996ck}, and
NS5-branes. A fruitful route for determining these corrections explicitly is
via a chain of non-perturbative symmetries and dualities \cite{RoblesLlana:2007ae,Saueressig:2007gi}.
A key role is played by the SL(2, $\IZ$) duality of the Type IIB
string \cite{Bohm:1999uk} and the symplectic covariance
of the Type IIA theory \cite{deWit:1996ix}, which are believed to hold also
non-perturbatively.\footnote{For a recent analysis of the interplay between these symmetries and
 wall-crossing phenomena see \cite{Manschot:2009ia}.} At the classical level,
the actions of these dualities on the physical fields parameterizing the moduli spaces can be
deduced from the dimensional reduction of the ten-dimensional supergravity
action. At the quantum level,
we then {\it define} the ``physical fields'' coordinatizing $\cM$ by the requirement that
they obey the classical
transformation laws with respect to SL(2, $\IZ$) on the Type IIB side and the symplectic group
on the Type IIA side, respectively.
Thus, by definition, the action of these transformations on $\cM$ does not receive quantum corrections.

Following this route, \cite{RoblesLlana:2006is} obtained exact results for the
D($-1$) and D1-instanton corrections
by implementing the SL(2, $\IZ$)-invariance of the Type IIB string. (For a recent rederivation
of some of these results from a topological string perspective see \cite{Collinucci:2009nv}.)
Using the classical limit of the mirror map together with the assumption
that the instanton numbers on the Type IIA and Type IIB side agree,
this result also yields the mirror symmetric
``A-type'' D2-instanton contributions \cite{Saueressig:2007dr,RoblesLlana:2007ae,Saueressig:2007gi}.
Subsequently, these corrections have been generalized to include all
D2-instantons \cite{Alexandrov:2008gh,Alexandrov:2009zh}
and there are also partial results on NS5-brane instantons
\cite{Alexandrov:2006hx,Pioline:2009qt}. Determining the exact contribution
of the latter will, however, require detailed knowledge of the
generalized mirror map, which provides a strong motivation for our investigation.

Our work heavily draws on
twistor space description of QK spaces \cite{Alexandrov:2008ds,Alexandrov:2008nk}.
This approach encodes the complicated quaternion-K\"ahler geometry
of $\cM$ in a set of simple holomorphic functions
appearing as transition functions between locally flat Darboux patches of the twistor space $\cZ$ of $\cM$.
These transition functions depend on the complex coordinates
on the twistor space, the so-called {\it twistor lines}, and play a similar role as
the holomorphic prepotential for the special K\"ahler spaces.
In this framework, the problem of determining the instanton corrections to $\cM$ is equivalent to
finding appropriate holomorphic functions which respect the non-perturbative symmetries
of the theory.

The strategy for constructing the generalized mirror map
is then to lift the non-perturbative duality transformations from $\cM$ 
to $\cZ$, where they induce transformations
of the twistor lines. The twistor lines then
transform in a particular representation of the duality group.
At the classical level, the actions of
SL(2,$\IZ$) $S$-duality and electric-magnetic rotations
have already been obtained in \cite{Alexandrov:2008gh}.
We will show that, locally,
these classical transformations remain uncorrected
upon including D-instanton contributions.
Imposing compatibility between the transformation laws of the physical fields and the twistor
lines then allows us to determine the instanton corrections to the generalized mirror map.
At the classical level, this computation has already been carried 
out in \cite{Alexandrov:2008gh} recovering the
result of \cite{Bohm:1999uk}.

Motivated by the observation that D1-instantons play an essential role in
smoothing the conifold singularity appearing in the perturbative
hypermultiplet metric \cite{Ooguri:1996me,Saueressig:2007dr},
we also investigate the effect of the D-instanton corrections
on the perturbative (1-loop) singularity of $\cM$ \cite{Robles-Llana:2006ez,Alexandrov:2007ec}.
In this context, we observe that the four-dimensional string coupling
(defined via the contact potential, see eq.\ \eqref{contact} below) does not become
strongly coupled,
while the moduli space metric contains a curvature singularity
at a finite value of the ten-dimensional string coupling.
The latter is not resolved by the instanton
corrections considered here, and we conjecture
that the resolution of this singularity will require the inclusion of NS5-branes.

The rest of the paper is organized as follows. In Sections 2 and 3 we review
the general twistor space construction of toric quaternion-K\"ahler manifolds
and its application to the perturbative hypermultiplet moduli spaces
arising in Type IIA and Type IIB string compactifications, respectively.
Section 4 contains our new description of the instanton corrected Type IIB twistor
space. In Appendices A and B we show that this formulation is gauge-equivalent to the
Type IIA description \cite{Alexandrov:2008gh}, establishing mirror symmetry at the level of the
twistor spaces. This result is used to derive the SL(2, $\IZ$)-transformation
of the instanton corrections to the twistor lines in the Type IIB framework and the non-perturbative
mirror map in Section 5. Section 6 contains our analysis of the perturbative singularities of $\cM$.
A brief discussion of our results together with an outlook is given in Section \ref{sec:7}.
Finally, the technical details about the SL(2, $\IZ$)-transformations of the Type IIB twistor lines
can be found in Appendix C.

\section{Twistor description of toric quaternion-K\"ahler spaces}
\label{sec_twist}

In this section we summarize the twistor space description
of (toric) QK manifolds recently developed in \cite{Alexandrov:2008nk}.
It is based on the Lebrun-Salamon theorem \cite{quatman,MR1001707,Alexandrov:2008nk},
which guarantees that the metric on $\cM$ can be recovered from the complex
contact structure\footnote{For a general introduction to contact geometry, see \cite{MR2194671}.}
on its twistor space $\cZ$.
Moving from $\cM$ to $\cZ$ brings the advantage
that all the geometrical data
can be encoded in a set of holomorphic transition functions
which relate the contact structure in different patches of $\cZ$.
These functions provide an invaluable tool when studying
deformations of QK spaces. In the following,
we will review these geometric objects
and their relation to the metric on $\cM$,
which will be central in our
discussion of instanton corrections to HM moduli spaces.

A QK manifold $\cM$ is a real $4d$-dimensional Riemannian manifold
with holonomy contained in $USp(d) \times SU(2)$. Locally it admits a triplet
of almost complex structures which satisfy the algebra of unit quaternions.
The holonomy constraint implies that $\cM$ carries a canonical SU(2)-connection
$\vec{p}$ with components $p_- = (p_+)^*$, $p_3 = (p_3)^*$,
which is given by the SU(2)-part of the Levi-Civita connection.
We call $\cM$ toric, if it admits $d+1$ commuting isometries.

The twistor space $\cZ$ is a $\CP$-bundle over $\cM$, i.e., $\cZ = \cM \times \CP$ locally,
and we use local coordinates $x^\mu$ on $\cM$ and the complex coordinate $\varpi$ on $\CP$,
respectively. $\cZ$ is K\"ahler-Einstein and its connection is given by $\vec{p}$.
 Furthermore, it carries a real structure $\tau$
which acts as the antipodal map on $\CP$ and leaves the coordinates of $\cM$
invariant,
$\tau\ :\ \{x^\mu,\varpi\} \mapsto \{x^\mu,-\bar\varpi^{-1}\}$.

The key ingredient in the construction is the complex contact
structure of $\cZ$. On the open covering $\hU_i$ of $\cZ$, the latter can be
represented by a set of holomorphic one-forms $\hCX\ui{i}$,
such that the holomorphic top form
\be\label{htf}
\kappa^{[i]}=
\hCX\ui{i} \wedge (\de\hCX\ui{i})^d \not = 0
\ee
is nowhere vanishing.
On each patch,
\be
\label{contact}
\hCX\ui{i} = 2\, \e^{\Phi\di{i}} \frac{D\varpi}{\varpi}\, ,
\ee
is proportional to the canonical $(1,0)$ form
$D\varpi =  \de\varpi + p_+ -\I p_3 \,\varpi + p_-\, \varpi^2$ and subject to the reality constraint
$\overline{\tau(\hCX\ui{i})}= - \hCX\ui{\bi}$ where
we assume that $\tau$ maps the patch $\hU_i$ onto $\hU_{\bi}$, which can always be
achieved through a refinement of the covering.
The ``contact potential'' $\Phi\di{i}\equiv\Phi\di{i}(x^\mu,\varpi)$
is a function on $\hU_i\subset\cZ$.
It is holomorphic along the $\CP$ fiber, defined up to an additive holomorphic
function on $\hU_i$, and chosen such that the right-hand
side of \eqref{contact} is a holomorphic (i.e. $\bar\pa$-closed) one-form.
Furthermore, the reality constraint on $\hCX\ui{i}$ implies that
$
\overline{\tau(\Phi\di{i})}=\Phi\di{\bi}.
$

By a variation of Darboux's theorem, for an appropriate open covering,
in each patch $\hat{\cU}_i$ one can find complex coordinates
$\xii{i}^\Lambda, \txii{i}_\Lambda, \ai{i}$ ($\Lambda=0,\dots,d-1$)
such that the $\hCX\ui{i}$ takes the canonical form
\be
\CX\ui{i}  \equiv  \de \alpi{i}
+  \xii{i}^\Lambda \, \de \txii{i}_\Lambda.
\label{con1fo}
\ee
We choose these coordinates to satisfy
\be
\label{rexixit}
\overline{\tau(\xii{i}^\Lambda)} = \xii{\bar \imath}^\Lambda\, ,
\qquad
\overline{\tau(\txii{i}_\Lambda)} = -\txii{\bi}_\Lambda
\, ,
\qquad
\overline{\tau(\ai{i})} = -\ai{\bi}
\, .
\ee
The global structure of $\cZ$ can then be encoded in a set of holomorphic transition functions
relating the sets of coordinates on the overlap of two patches $\hat\cU_i \cap \hat\cU_j$,
thereby preserving \eqref{con1fo}. Together with
the $d+1$ additional real numbers $c_\Lambda, c_\alpha$,
called the ``anomalous dimensions'', these transition functions
contain all the geometric information of the twistor space and the corresponding QK base.

Extracting the metric on $\cM$ requires the construction of the twistor
lines of $\cZ$ by expressing the complex coordinates
$\xii{i}^\Lambda, \txii{i}_\Lambda, \ai{i}$ in terms of the coordinates $\{x^\mu, \varpi\}$.
For a generic QK metric, where the transition
functions depend on all coordinates $\xii{i}^\Lambda, \txii{i}_\Lambda, \ai{i}$,
this is fairly difficult. For the purpose of this paper it suffices, however, to consider the case where
$\cM$ is toric. In this case
 the moment maps associated to the $d+1$
isometries \cite{MR872143} provide
independent global $\cO(2)$ sections on $\cZ$, which can
be taken to be the complex coordinates $\xii{i}^\Lambda$ and the unit function.
The resulting twistor lines $\xii{i}^\Lambda$ are globally defined and take the form
\be
\label{gxi}
 \xi^\Lambda \equiv \xii{i}^\Lambda = Y^\Lambda \varpi^{-1} + A^\Lambda - \bY^\Lambda \varpi\,
\ee
on all patches $\hU_i$. In fact, one can choose $Y^0\equiv \cR$ to be real
by fixing the $U(1)$-action corresponding to phase rotations of $\varpi$.
Together with the $d+1$ additional real coordinates $\{ B_\Lambda, B_\alpha\}$ introduced in \eqref{eqmuqh},
$\cR, Y^a, \bY^a, A^\Lambda$ provide a convenient coordinate system on $\cM$.

As a consequence of \eqref{gxi}, the complex coordinates
$\xi^\Lambda, \txi_\Lambda$ and  $\alpha$
must now be related by transition functions which preserve $\xi^\Lambda$
and the unit function. The general form of such contact transformations is given by
\bea
\label{contactmu}
\txii{i}_\Lambda &=&  \txii{j}_\Lambda - \pa_{\xi^\Lambda} \hHij{ij}\, ,
\qquad \ai{i} =  \ai{j} - \hHij{ij}
  +\xi^\Lambda\pa_{\xi^\Lambda} \hHij{ij} \, ,
\eea
where the transition functions $\hHij{ij}(\xi)$ are independent of $\txii{i}_\Lambda, \ai{i}$.
The reality condition \eqref{rexixit} and the
consistency conditions appearing on the overlap of three patches
furthermore imply the additional constraints
\be
\label{consisth}
\overline{\tau(\hHij{ij})} = -\hHij{\bi\bj}\, ,
\qquad
\hHij{ij}+ \hHij{jk} = \hHij{ik}\, .
\ee
Besides, it is important to note that
the transition functions $\hHij{ij}$ do not specify the twistor space uniquely,
but are subject to the gauge equivalence
\be\label{gaugeeq}
\hHij{ij} \mapsto  \hHij{ij}+ T^{[i]}-T^{[j]}\, ,
\ee
where $T^{[i]}(\xi^\Lambda)$ are holomorphic functions regular in the patch $\hU_i$.
Essentially they capture
the possibility to perform a local change of coordinates in $\hat{\cU}_i$, which
leaves the contact form \eqref{con1fo} invariant.
We shall often abuse notation and define $\hHij{ij}$ away from the overlap $\hU_i \cap \hU_j$
(in particular when the two patches do not intersect) using analytic continuation and the
second equation in \eqref{consisth} to interpolate from $\hU_i$ to $\hU_j$. Ambiguities in the
choice of path can be dealt with on a case by case basis.

The gluing conditions \eqref{contactmu} together with the requirement
that the contact form takes the form \eqref{contact}
are sufficient to determine the twistor lines $\txii{i}_\Lambda, \ai{i}$ for the toric case
\cite{Alexandrov:2008nk}:\footnote{Formulas for the twistor lines
arising from an infinitesimal deformation away from the toric case
have been obtained in \cite{Alexandrov:2008nk}.}
\bea
 \label{eqmuqh}
\txii{i}_\Lambda &=&
\frac{\I}{2}\, B_\Lambda+\frac12\sum_j\oint_{C_j}\frac{\de\varpi'}{2\pi\I\varpi'}\,
\frac{\varpi'+\varpi}{\varpi'-\varpi}\, \pa_{\xi^\Lambda}  \hHij{ij}(\xi (\varpi'))
+\cij{+}_\Lambda \log \varpi\, ,
\\
\ai{i} &=& \frac{\I}{2}\, B_\alpha+\frac12\sum_j\oint_{C_j}\frac{\de\varpi'}{2\pi\I \varpi'}\,
\frac{\varpi'+\varpi}{\varpi'-\varpi}\, \left[ H - \xi^\Lambda  \pa_{\xi^\Lambda} H
 \right]^{[ij]}
+\cij{+} _\alpha \log \varpi
+\cij{+}_\Lambda\(Y^\Lambda \varpi^{-1} + \bY^\Lambda \varpi\) \, .
\nn
\eea
Here, $\varpi\in \cU_i$, with $\cU_i$
denoting the projection of $\hat{\cU}_i$ to $\CP$, and $C_j$ is a contour surrounding $\cU_j$.
The ``integration constants'' $B_\Lambda, B_\alpha$
provide the extra $d+1$ coordinates on $\cM$ mentioned above.
Note that the twistor lines are not regular in the patches around $\varpi=0$ and $\varpi=\infty$
but may contain singular terms.
These terms are weighted by the anomalous dimensions $c_\Lambda, c_\alpha$ which control the singular behavior
at these points. These are the only singularities admissible
by regular metrics.
Moreover, the contact potential turns out to be independent of $\varpi$ and the same in all patches.
Explicitly it is given by
\be
\label{eqchi}
e^{\Phi}=\frac14 \, \sum_j\oint_{C_j}\frac{\de\varpi'}{2\pi\I \varpi'}
\(\varpi'^{-1} Y^{\Lambda}-\varpi' \bY^{\Lambda} \)\pa_{\xi^\Lambda}
{\hHij{ij}}(\xi(\varpi')) +\frac12\( \cij{+}_\Lambda  A^\Lambda +  \cij{+}_\alpha\) \, .
\ee
Note that due to consistency conditions \eqref{consisth}, the index $i$ of
the transition functions in
\eqref{eqmuqh}, \eqref{eqchi} can be replaced by any other patch index without affecting the result.

Given the twistor lines \eqref{gxi} and \eqref{eqmuqh}, the QK metric on $\cM$ can be obtained
using the general formalism \cite{Alexandrov:2008nk}, which considerably simplifies for the toric case.
This formalism utilizes the Laurent expansion of the twistor
lines in the patch $\hat{\cU}_+$ around $\varpi = 0$.
Substituting this expansion into the contact one-form \eqref{con1fo} and
comparing to \eqref{contact} allows to determine the SU(2) connection $\vec p$ and
the contact potential in terms of the Laurent coefficients of this expansion.
The SU(2) connection then gives the triplet
of quaternionic forms $\vec{\omega}_{\cM}$, and in particular
\be
\label{om3p}
\omega_{\cM,3} = \de p_3 +2\I \, p_+ \wedge p_-  \, .
\ee
Upon determining the almost complex structure $J_3$, the metric then follows
from $g_{\cM}(J_3 X, Y) = \omega_{\cM,3}(X,Y)$. The former can be specified
through a basis of local one-forms on $\cM$ of Dolbeault type (1,0) with respect to $J_3$.
Let us denote by $\txi^{[+]}_{\Lambda,0}$ and $\alpha^{[+]}_0$ the constant terms in the Laurent
expansion of $\txii{+}_\Lambda$ and $\ai{+}$, respectively.
Then by expanding the holomorphic one-forms
$\rmd \xi^\Lambda, \rmd \txi_\Lambda^{[+]}$, and $\rmd \alpha^{[+]}$ around $\varpi=0$
and projecting the result along the base $\cM$, it can be shown that a suitable basis is given
by\footnote{Up to overall factors, these correspond to the $(1,0)$-forms $\Pi^a, \tilde{\Pi}_I$
introduced in \cite{Alexandrov:2008nk}, which are further simplified
by utilizing that $p_+$ itself is of Dolbeault-type $(1,0)$.}
\be
\label{defPi}
\Pi^a = \de \( Y^a/\cR\),
\quad
\tilde{\Pi}_\Lambda = \de \txi^{[+]}_{\Lambda,0}+c_\Lambda\de \log\cR,
\quad
\tilde{\Pi}_\alpha = -\de \alpha^{[+]}_0 +c_\Lambda \de A^\Lambda -c_\alpha\de\log\cR,
\ee
where $a$ runs over $1,\ldots,d-1$.
These one-forms will play a central role in our discussion
of the regularity of the metric on $\cM$ in Section \ref{sec_reg}. In this context, it is often
 convenient to trade the coordinate $\cR$ for the variable $\e^\Phi$.
As we shall see below, this is natural for the hypermultiplet moduli space,
since the contact potential $\e^\Phi$ is identified with the four-dimensional dilaton $\e^\phi$.

\section{Twistor description of perturbative HM moduli spaces}
\label{sec_perturb}
We now review the twistor space description of
 the perturbative hypermultiplet moduli spaces arising
from compactifying Type II string theory on a CY threefold.
This will provide our starting point for determining the instanton corrections to the classical
mirror map in Section \ref{sec_mirror}. Owed to the gauge-invariance
of the $p$-forms appearing in the original ten-dimensional
action before compactification, the perturbative hypermultiplet moduli spaces admit
a Heisenberg group of isometries and therefore fall into the
class of toric QK manifolds discussed in the previous section.

\subsection{Type IIA compactified on a CY threefold $X$}
\label{subsec_iia}

The hypermultiplet moduli space $\cM_{C}(X)$ in Type IIA string theory
compactified on a CY
threefold $X$ is a QK manifold of quaternionic
dimension $d=h^{2,1}(X)+1$ \cite{Cecotti:1988qn,Ferrara:1989ik,Bodner:1990zm}.
It describes the dynamics of the complex structure moduli
$X^\Lambda=\int_{\gamma^\Lambda} \Omega$,
$F_\Lambda=\int_{\gamma_\Lambda} \Omega$, the RR scalars
$\zeta^\Lambda,\tzeta_\Lambda$ originating as similar integrals of the RR 3-form,
the four-dimensional dilaton
$\e^{\phi}=1/g_{(4)}^2$ and the Neveu-Schwarz (NS) axion $\sigma$, dual to the
NS two-form $B$ in four dimensions.
Here, $\{ \gamma^\Lambda, \gamma_\Lambda\}$
represent a symplectic basis of A- and B-cycles in $H_3(X,\IZ)$
with intersection product $\langle \gamma^\Lambda, \gamma_\Sigma\rangle=\delta^\Lambda_\Sigma$. The sets
$\{X^\Lambda, F_\Lambda(X)\}$ and $\{\zeta^\Lambda, \tzeta_\Lambda \}$ transform as symplectic vectors
with respect to electric-magnetic duality Sp($2d$, $\IZ$).
The $X^\Lambda$ provide
a set of homogeneous coordinates on the space of complex structure deformations $\cK_{C}(X)$,
and (away from the vanishing locus of $X^0$) may be traded for
the inhomogeneous coordinates $z^a=X^a/X^0$.

At string tree level, $\cM_{C}(X)$ can be obtained from the special K\"ahler space
$\cK_{C}(X)$ (describing the vector multiplet moduli space
of Type IIB strings compactified on the same CY $X$) by
the $c$-map \cite{Cecotti:1988qn,Ferrara:1989ik}.
The latter space is completely characterized by
the prepotential $F(X^\Lambda)$,
a homogeneous function of degree two of the A-type periods $X^\Lambda$, such that
the B-type periods are given by $F_\Lambda=\pa F/\pa X^\Lambda$.
Therefore, the same is true for the tree level HM moduli space.
At one-loop, the metric on $\cM_{C}(X)$ receives a correction proportional to the
Euler class $\chi_X=2(h^{1,1}(X)-h^{2,1}(X))$.
Based on the
string theory amplitudes \cite{Antoniadis:1997eg,Gunther:1998sc},
the one-loop corrected QK metric was calculated
in \cite{Robles-Llana:2006ez}.
It is believed to be the correct metric on $\cM_{C}(X)$ to all orders in perturbation
theory \cite{Gunther:1998sc,Robles-Llana:2006ez,Alexandrov:2007ec,Alexandrov:2008nk}.

The twistor space formulation of $\cM_{C}(X)$ was worked out in \cite{Alexandrov:2008nk}.
As illustrated in the left diagram of Fig.\ \ref{Fig.2}, it
utilizes two patches $\hU_+$, $\hU_-$
which project to open disks centered around $\varpi=0$ and $\varpi=\infty$ on $\CP$,
and a third patch $\hU_0$ which projects to the rest of $\CP$. The transition functions
between these patches and the anomalous dimensions are given by
\be
\label{symp-cmap}
\hHij{0+}= -\frac{\I}{2}\, F(\xi^\Lambda)\, ,
\qquad
\hHij{0-}=-\frac{\I}{2}\,\bF(\xi^\Lambda)\, ,
\qquad
\ci{+}_\alpha = \frac{\chi_X}{96\pi} \, ,
\ee
with $\ci{\pm}_\Lambda=0$. The non-vanishing $\ci{\pm}_\alpha$ incorporates the one-loop correction.

The twistor lines arising from \eqref{symp-cmap} are
readily computed from \eqref{eqmuqh}. Motivated by symplectic invariance,
it is convenient to express the twistor line $\alpha^{[i]}$ in terms of
\be
\label{defrho}
\aA^{[i]} \equiv 4\I  \alpha^{[i]} + 2 \I \txi_\Lambda^{[i]} \xi^\Lambda \, ,
\ee
so that the symplectic form \eqref{con1fo} becomes
\be\label{symp2}
\CX^{[i]} = \frac{1}{4\I}\, {\rmd} \aA^{[i]} +
\frac{1}{2} \( \xi^\Lambda {\rmd} \txi^{[i]}_\Lambda - \txi^{[i]}_\Lambda {\rmd} \xi^\Lambda \) \, .
\ee
The twistor lines in the patch $\hU_0$ resulting from \eqref{symp-cmap} are then
given by \cite{Neitzke:2007ke,Alexandrov:2008nk}
\be
\label{gentwi}
\begin{array}{rcl}
\xi^\Lambda &=& \zeta^\Lambda + \cR
\left( \varpi^{-1} z^{\Lambda} -\varpi \,\bz^{\Lambda}  \right)\, ,
\\
-2 \I \txii{0}_\Lambda &=& \tzeta_\Lambda + \cR
\left( \varpi^{-1} F_\Lambda(z)-\varpi \,\bF_\Lambda(\bz) \right)\, ,
\\
\aA^{[0]} &=& \sigma + \cR
\left(\varpi^{-1} W(z)-\varpi \,\bar W(\bz) \right) +\frac{\I\chi_X}{24\pi} \,\log \varpi \, .
\end{array}
\ee
Here we set
\be
\label{defW}
W(z) \equiv  F_\Lambda(z) \zeta^\Lambda - z^\Lambda \tzeta_\Lambda\, ,
\ee
and used the relation between the generic coordinates $Y^a,A^\Lambda, B_\Lambda, B_\alpha$ introduced
in the previous section and the physical Type IIA fields
\be
\zeta^\Lambda=A^\Lambda\, ,
\quad
\tzeta_\Lambda=B_{\Lambda}+A^\Sigma \Re F_{\Lambda\Sigma}(z)\, ,
\quad
\sigma =- 2 B_\alpha  - A^\Lambda B_\Lambda\, ,
\quad
Y^a = \cR\, z^a \, .
\label{relABzeta}
\ee
The contact potential resulting from \eqref{symp-cmap} is given by
\be\label{phipertB}
\e^{\Phi_{\rm pert}}= \frac{\cR^2}{4}\,K(z,\bar{z})+\frac{\chi_X}{192\pi} \, ,
\ee
where $K(z,\bz)\equiv-2 \Im(\bz^\Lambda F_\Lambda)$.
This Type IIA twistor space formulation is adapted to the
symplectic covariance of the theory. Indeed, 
$\{\xi^\Lambda,-2 \I \txii{0}_\Lambda\}$ transforms as a symplectic
vector while $\aA^{[0]}$ and $\e^{\Phi_{\rm pert}}$ are symplectic invariants. 
Thus the contact form \eqref{symp2} is also invariant, making the symplectic 
covariance manifest.
This completes the twistor description of the perturbative
Type IIA HM moduli space.

\subsection{Type IIB compactified on a CY threefold $Y$}
\label{subsec_iib}
%
The QK manifold $\cM_K(Y)$ arising from compactifying
Type IIB string theory on a CY threefold $Y$ has
quaternionic dimension $d=h^{1,1}(Y)+1$.
It describes the dynamics of the
K\"ahler moduli $z^a\equiv b^a + \I t^a=\int_{\gamma^a} \mathcal{J}$,
the RR scalars
$c^0,c^a,c_a,c_0$ obtained by integrating appropriate combinations of RR forms along
even dimensional cycles,
the four-dimensional dilaton $\e^\phi$ and the NS axion $\psi$.
Here $\mathcal{J}\equiv B+\I\, J=z^a \omega_a$ is the
complexified \kahler form on $Y$. Furthermore,  $\gamma^a$,
$a=1,\dots, h^{1,1}(Y)$, form a basis of 2-cycles (Poincar\'e dual to 4-forms $\omega^a$),
$\gamma_a$ a basis of
4-cycles (Poincar\'e dual to 2-forms $\omega_a$),
and $ \kappa_{abc}=\int_Y \omega_a \omega_b \omega_c
=\langle \gamma_a, \gamma_b, \gamma_c\rangle$
is the triple intersection product in $H_4(Y,\IZ)$.
In the large volume limit, the four-dimensional dilaton $\phi$ is related to the ten-dimensional
string coupling $g_{(10)}$ via $\e^{\phi}=\hf\,V(t^a)/(g_{(10)})^2$, where
$V(t^a)=\frac{1}{6}\int_Y J\wedge J\wedge J = \frac16\, \kappa_{abc}t^a t^b t^c$
is the volume of $Y$ in string units. The  ten-dimensional
coupling $\tau_2\equiv 1/g_{(10)}$ and the RR axion $\tau_1\equiv c^0$
can be combined into the ten-dimensional axio-dilaton field $\tau = \tau_1+\I \tau_2$.

At string tree level, the metric on $\cM_K(Y)$ can be obtained from the special \kahler space
$\cK_{K}(Y)$
via the $c$-map and thus is determined by the prepotential $F(X^\Lambda)$. The latter
receives world-sheet instanton corrections which can conveniently be found via the
classical mirror map \cite{Candelas:1990rm,Hori:2003ic}. Its large volume expansion takes
the form
\be
\label{lve}
F(X^\Lambda)=-\kappa_{abc} \,\frac{X^a X^b X^c}{6 X^0} + \chi_Y\,
\frac{\zeta(3)(X^0)^2}{2(2\pi\I)^3}
-\frac{(X^0)^2}{(2\pi\I)^3}\sum_{\hat\gamma_+} n_{q_a}^{(0)}\, \Li_3\left(
\e^{2\pi \I  q_a X^a/X^0}\right)\, ,
\ee
where $\hat\gamma$ denotes the set of charges $\{q_a\}$ and its subset $\hat\gamma_+$
corresponds to effective homology classes
$q_a\gamma^a\in H_2^+(Y)$ (i.e., $q_a\geq 0$ for all $a$, not all of them vanishing simultaneously).
Furthermore, $n_{q_a}^{(0)}$ is the genus zero BPS invariant in the homology
class $q_a \gamma^a$,
$\Li_s(x)=\sum_{m=1}^\infty m^{-s}x^m$ is the polylogarithm function,
and $\chi_Y$ is the Euler number of $Y$. Note that the last two terms
in \eqref{lve} may be combined by including the zero class
$q_a=0$ into the sum and setting $n_0^{(0)}=-\chi_Y/2$.
Similarly to the Type IIA case,
the $c$-map metric on $\cM_K(Y)$ receives a one-loop correction proportional to the Euler
class $\chi_Y$.

Classically, i.e., at string tree-level and leading order in the $\alpha'$ expansion,
$\cM_K(Y)$ admits an isometry group SL(2, $\IR)$,
acting on the physical fields as \cite{Gunther:1998sc,Bohm:1999uk}
\be\label{SL2Z}
\begin{split}
&\quad \tau \mapsto \frac{a \tau +b}{c \tau + d} \, ,
\qquad
t^a \mapsto t^a |c\tau+d| \, ,
\qquad
c_a\mapsto c_a\, ,
\\
&
\begin{pmatrix} c^a \\ b^a \end{pmatrix} \mapsto
\begin{pmatrix} a & b \\ c & d  \end{pmatrix}
\begin{pmatrix} c^a \\ b^a \end{pmatrix}\, ,
\qquad
\begin{pmatrix} c_0 \\ \psi \end{pmatrix} \mapsto
\begin{pmatrix} d & -c \\ -b & a  \end{pmatrix}
\begin{pmatrix} c_0 \\ \psi \end{pmatrix} \, ,
\end{split}
\ee
with $ad-bc=1$, which is inherited from the SL(2, $\IR)$ invariance
of the ten-dimensional Type IIB supergravity action. The $\alpha'$- and $g_s$-corrections
break this symmetry to the discrete group SL(2, $\IZ)$, which is expected to be
a symmetry of the full quantum theory.

The twistor space $\cZ$ of $\cM_K(Y)$ has the same patch structure as its Type IIA
counterpart, cf.\ Fig.\ \ref{Fig.1}. Provided that $Y=\tX$ is the mirror CY of $X$,
it also employs the same transition functions
\eqref{symp-cmap}. Thus, also the twistor lines underlying the perturbative $\cM_K(\tX)$
and $\cM_C(X)$ are identical.
Using the large volume expansion \eqref{lve}
and identifying $\cR=\tau_2/2$ (as will become clear in \eqref{physmap} below), the contact potential
\eqref{phipertB} has the large volume expansion
\be
\label{phipertBlv}
\begin{split}
\e^{\Phi_{\rm pert}}
& = \frac{\tau_2^2}{2} \,V(t^a)-\frac{\chi_Y\zeta(3)}{8(2\pi)^3}\,\tau_2^2
+ \e^{\Phi_{\rm ws}}- \frac{\chi_Y}{192\pi}\, ,
\end{split}
\ee
where
\be
\label{phiws}
\e^{\Phi_{\rm ws}} =\frac{\tau_2^2}{4(2\pi)^3}\sum_{\hat\gamma_+} n_{q_a}^{(0)}
\Re\left[ \Li_3 \left( \e^{2\pi \I q_a z^a} \right) + 2\pi q_a t^a\,
\Li_2 \left( \e^{2\pi \I q_a z^a} \right)  \right]
\ee
is the world-sheet instanton contribution.
In the large volume limit, $\e^{\Phi_{\rm pert}}$ coincides with the four-dimensional dilaton $\phi$,
and may in fact be adopted as its definition in the quantum regime.

The classical SL(2, $\IR)$-invariance \eqref{SL2Z} can be lifted to the twistor space. In this context, it
is useful to adapt the twistor line $\alpha^{[i]}$ to the SL(2, $\IR)$ symmetry and work with
\be
\aB = -\frac{\I}{4}\,\aA+\hf\,\txi_\Lambda\xi^\Lambda=\alpha+\txi_\Lambda\xi^\Lambda \, .
\label{defha}
\ee
Then the SL(2, $\IR)$ action on the complex coordinates on the patch
$\hU_0$ takes the form \cite{Alexandrov:2008gh}
\be
\label{SL2Zxi}
\begin{split}
\xi^0 &\mapsto \frac{a \xi^0 +b}{c \xi^0 + d} \, , \qquad
\xi^a \mapsto \frac{\xi^a}{c\xi^0+d} \, , \qquad
\txi_a \mapsto \txi_a +  \frac{\I\, c}{4(c \xi^0+d)} \kappa_{abc} \xi^b \xi^c\, ,
\\
\txi_0 &\mapsto   (c\xi^0+d)\txi_0- c \, \aB +c\xi^a\txi_a
+\frac{\I c^2}{12}\,\frac{\kappa_{abc}\xi^a\xi^b\xi^c}{c\xi^0+d} \, , \\
\aB &\mapsto
\frac{\aB}{c\xi^0+d}+\frac{\I c}{12}\,\frac{\kappa_{abc}\xi^a\xi^b\xi^c}{(c\xi^0+d)^2}.
\end{split}
\ee
Under the action \eqref{SL2Zxi},  the complex
contact one-form transforms by an overall holomorphic factor
$\hCX\ui{i}\to \hCX\ui{i}/(c\xi^0+d)$ so that the complex
contact structure remains invariant.

The holomorphic contact action \eqref{SL2Zxi} on $\cZ$ decomposes into the isometric action
\eqref{SL2Z} on $\cM_K(Y)$ and
a SU$(2)$ rotation on the fiber. The latter is given by the following transformation
of the fiber coordinate $\varpi$:
\be
\varpi \mapsto  \frac{c \tau_2 + \varpi  (c \tau_1 + d) +
\varpi |c \tau + d| }{(c \tau_1 + d) + |c \tau + d| - \varpi c \tau_2}\, .
\label{transz}
\ee
The SL(2, $\IR)$ transformation of the complex coordinates
$\xi^\Lambda, \txi_\Lambda^{[0]}, \alpha_{\rm B}^{[0]}$, given by eq.\ \eqref{SL2Zxi},
should be consistent with the transformation of the (classical part of the)
twistor lines \eqref{gentwi} induced by the transformation laws \eqref{SL2Z} and \eqref{transz}.
This condition allows to determine the classical relation between the
physical Type IIA fields $(\cR,z^a,\zeta^\Lambda,\tzeta_\Lambda,\sigma)$ and
their Type IIB counterparts $(\tau, b^a, t^a, c^a,$ $ c_a,c_0,\psi)$ \cite{Alexandrov:2008gh}
\be
\label{physmap}
\begin{split}
\cR& =\frac12\,\tau_2\, , \qquad
Y^a=\frac12\, \tau_2 \, z^a\, ,\qquad
\zeta^0=\tau_1\, ,
\qquad
\zeta^a = - (c^a - \tau_1 b^a)\, , \\
\tzeta_a &=  c_a+ \frac{1}{2}\, \kappa_{abc} \,b^b (c^c - \tau_1 b^c)\, ,
\qquad
\tzeta_0 =\, c_0-\frac{1}{6}\, \kappa_{abc} \,b^a b^b (c^c-\tau_1 b^c)\, ,
\\
\sigma &= -2 (\psi+\frac12  \tau_1 c_0) + c_a (c^a - \tau_1 b^a)
-\frac{1}{6}\,\kappa_{abc} \, b^a c^b (c^c - \tau_1 b^c)\, .
\end{split}
\ee
This relation constitutes the classical limit of the generalized mirror map
and agrees with the identification found via the dimensional reduction of
the classical ten-dimensional Type IIB supergravity action on $Y$ \cite{Gunther:1998sc,Bohm:1999uk}.
The explicit construction of the quantum version of this map, including
the perturbative as well as the worldsheet, D($-1)$ and D1-instanton corrections,
will be the subject of Section \ref{sec_mirror}.

\section{Instanton corrected Type IIB HM moduli space}
\label{sec_instA}

We now proceed by dressing up the perturbative Type IIB twistor
space description of the last subsection by
including D($-$1) and D1-instanton corrections.
In this case $\cM_K(Y)$ still possesses $d+1$ commuting
isometries\footnote{These are broken once D3, D5 and NS5-brane corrections are included.}
and thus falls into the class of toric QK spaces discussed in Section \ref{sec_twist}.
The corresponding instanton corrections were found by carrying out an
SL(2, $\IZ)$ completion of the so-called ``tensor potential" \cite{deWit:2006gn},
which, up to the scale factor,  corresponds to the contact potential \eqref{eqchi}
in the twistor construction. Here we provide the complete description of the
instanton corrected twistor space including the transition functions and the twistor lines.
The gauge equivalence between this novel formulation and the Type IIA twistor space description
 \cite{Alexandrov:2008gh} is established in Appendix \ref{sect:3}
 with more computational details relegated to Appendix \ref{sec_Pois}.

Adapting the results \cite{RoblesLlana:2006is}, the instanton corrected Type IIB contact
potential is expressed in terms of a generalized Eisenstein series
\be
\e^{\Phi_{\rm IIB}} = \frac{\tau_2^2}{2} \,V(t^a)
+\frac{\sqrt{\tau_2}}{8(2\pi)^3}\sum_{q_a\geq 0} n_{q_a}^{(0)}
{\sum\limits_{m,n}}'\frac{\tau_2^{3/2}}{|m\tau+n|^3}\(1+2\pi |m\tau+n|q_a t^a\)\e^{-S_{m,n, q_a}} \, .
\label{phiinv}
\ee
Here the prime indicates that the $(m,n)$-sum excludes $(0,0)$.
Furthermore,
\be
S_{m,n,q_a} = 2 \pi q_a (| m \tau + n| t^a - \I m c^a - \I n b^a)
\ee
is the classical action of a $(p,q)$-string
(or rather $(m,n)$-string) wrapped on the 2-cycle $q_a \gamma^a\in H_2(Y,\IZ)$ and
the $n_{q_a}^{(0)}$ are the BPS invariants
introduced in \eqref{lve}.
The contribution of the D($-1$)-instantons
is recovered as the ($q_a$=0)-sector of the sum.
Observe that the contact potential is actually
independent of the RR fields $c_a, c_0$ and the NS axion $\psi$. With respect to the
SL(2, $\IZ)$-transformations \eqref{SL2Z} it transforms as
a modular form $\e^{\Phi_{\rm IIB}} \mapsto \e^{\Phi_{\rm IIB}}/|c\tau+d|$.

The projective superspace description  \cite{RoblesLlana:2007ae} suggests that the instanton
terms in \eqref{phiinv} can be captured by the holomorphic
function\footnote{Here and in the following, for $q_a=0$ the sum over $n$ is defined
by first summing the contributions of $n$ and $-n$ which leads to the asymptotics $\sim n^{-2}$.}
\be
\label{prepotGG}
\begin{split}
G_{\rm IIB}(\xi)&=  -\frac{\I}{(2\pi)^3}\sum\limits_{q_a\ge 0} n_{q_a}^{(0)}
\sum\limits_{n\in \IZ \atop m>0}\frac{e^{-2\pi \I m q_a\xi^a}}{m^2(m\xi^0+n)} ,
\end{split}
\ee
with $\xi^\Lambda(\varpi)$ given by \eqref{gentwi}.
This function has a dense set of poles on the real axis
\be
\varpi_\pm^{m,n} = \frac{ m \tau_1 + n \mp | m\tau + n |}{m \tau_2}\, ,
\qquad
\varpi^{m,n}_+ \varpi^{m,n}_- = -1,
\label{poles}
\ee
corresponding to the zeros of $m\xi^0 + n$ expressed in terms of Type IIB fields
using \eqref{physmap} (see \eqref{IIBtwi} below).
For $m>0$, the poles satisfy $\varpi^{m,n}_+<0$ and $\varpi^{m,n}_->0$, respectively.
Each pole captures the contribution of a particular $(m,n)$-instanton configuration.
This can easily be seen by noting that 
the residue of $G_{\rm IIB}$ at $\varpi^{m,n}_\pm$ gives rise to 
exponential terms containing the corresponding instanton action.
The ``off-shell'' $G_{\rm IIB}$
does not transform as a modular form under SL(2, $\IZ)$, however.

\begin{figure}[t]
\renewcommand{\baselinestretch}{1}
\begin{center}
\leavevmode
\put(-215,6){\includegraphics[width=0.39\textwidth]{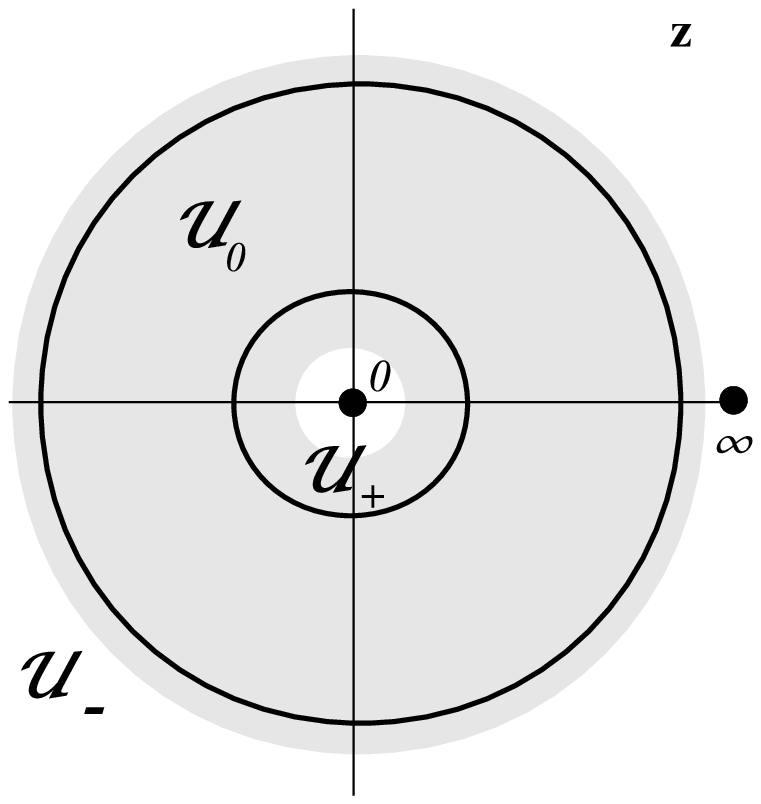}}
\hspace{5mm}
\leavevmode
\put(-30,95){\includegraphics[width=0.10\textwidth]{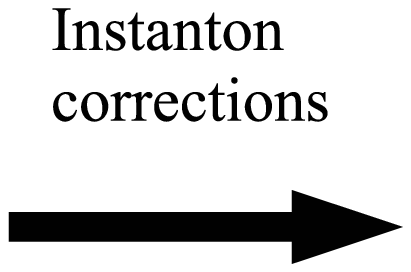}}
\leavevmode
\put(27,4){\includegraphics[width=0.47\textwidth]{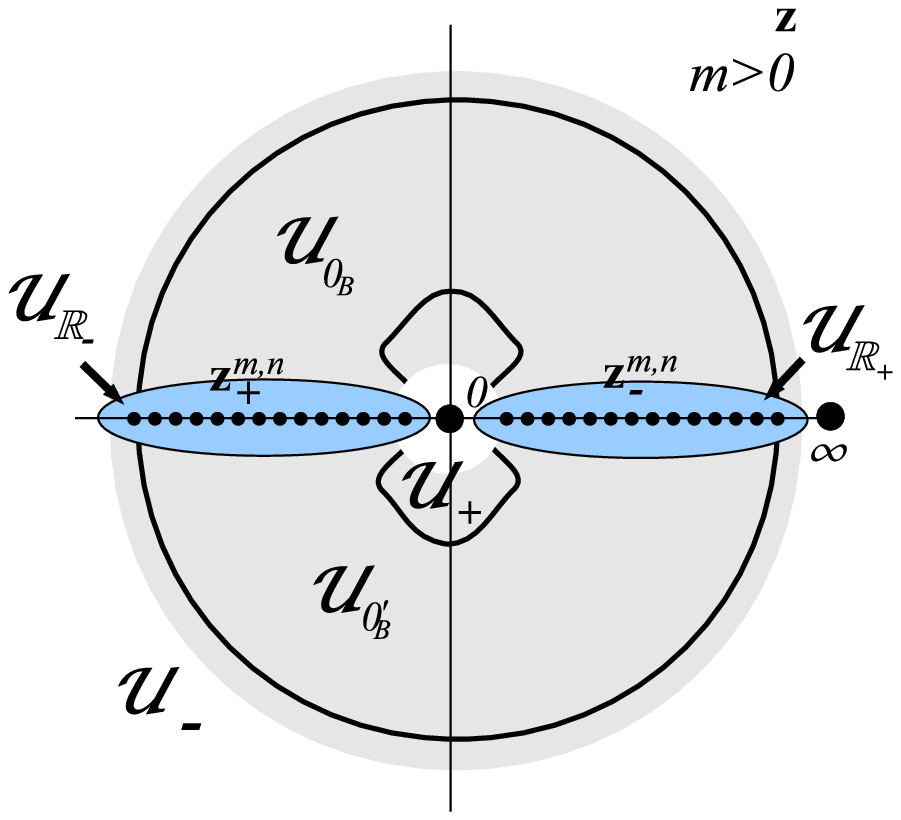}}
\end{center}
\parbox[c]{\textwidth}{\caption{\label{Fig.1}{Transition from
the perturbative to the instanton corrected Type IIB twistor space.
The corrections are encoded in the poles $\varpi^{m,n}_\pm$
located in the new patches $\cU_{\IR_\pm}$. This should be compared
to the gauge-equivalent Type IIA twistor space shown in Fig.\ 2,
which encodes the instanton corrections in two BPS rays along the
imaginary axis.}}}
\end{figure}

The inclusion of the instanton corrections in the twistor space description
is then depicted in Fig.\ \ref{Fig.1}.
The $\CP$ is covered by six patches:
the two (classical) patches $\cU_\pm$ surround the north and south
poles, two patches $\cU_{\bpm}$ encircle the negative and positive real axis and contain
all poles $\varpi_\pm^{m,n}$, and the remaining two patches $\cU_{\ob}$, $\cU_{\opb}$
cover the upper and lower half-planes of the $\varpi$-plane, respectively.\footnote{Strictly speaking,
this description is not satisfactory
because the poles are dense on the real axis and, in particular, accumulate near $\varpi=0$.
To make the construction rigorous,
one should cut off the sum over $m,n$ in \eqref{prepotGG} to  $m\leq M,\ |n|\leq N$.
This defines a regularized twistor space,
which converges to the desired space in the limit $M,N\to\infty$.
\label{foot_cover}}
The transition functions on the overlaps including $\cU_{0_B}$ are
\be
\label{symp-inst}
\begin{split}
\hHij{\ob +}= -\frac{\I}{2}\, F(\xi^\Lambda)\, ,
& \qquad
\hHij{\ob -}=-\frac{\I}{2}\,\bF(\xi^\Lambda)\, ,
\\
\hHij{\ob \bp}=-\frac{\I}{2}\, G_{\rm IIB}(\xi^\Lambda)\, ,
& \qquad
\hHij{\ob \bm}=-\frac{\I}{2}\,\bG_{\rm IIB}(\xi^\Lambda)\, ,
\end{split}
\ee
The ones on the overlap with $\cU_{\opb}$ are given by the same expressions 
and follow from replacing $0_B \mapsto \opb$ in \eqref{symp-inst}.
The anomalous dimensions $c_\Lambda, c_\alpha$ vanish identically since
the perturbative one-loop contribution $c_\alpha$, eq.\ \eqref{symp-cmap},
is now incorporated as a part of the $(q_a = 0)$-sector of the sum \eqref{prepotGG}.
(An explanation of how this works can be found in Appendix \ref{ssect:gt}.)
This data completely specifies the twistor space underlying the Type IIB HM moduli space.

Our next task is to compute the Type IIB twistor lines, including the instanton corrections, by expressing
$\xi^\Lambda, \txi_\Lambda, \aB$ in terms of the physical Type IIB fields and the fiber coordinate $\varpi$.
The twistor line $\xi^\Lambda$ can be obtained from \eqref{gxi} by first going to
the Type IIA variables via \eqref{relABzeta} and subsequently applying the 
mirror map \eqref{physmap}
\be\label{IIBtwi}
\xi^0 =  \tau_1 + \frac{\tau_2}{2} \( \varpi^{-1} - \varpi \) \, , \qquad
\xi^a =  - \( c^a - \tau_1 b^a \) + \frac{\tau_2}{2} \( z^a \, \varpi^{-1} - \bar{z}^a \, \varpi \) \, .
\ee
The off-shell formulation \cite{Saueressig:2007dr,RoblesLlana:2007ae} thereby guarantees, that the relation between
the physical fields entering into $\xi^\Lambda$ does not receive
quantum corrections from D($-1)$ and D1-instantons. As a
crosscheck, one can also compute the contact potential
\eqref{eqchi} from the transition functions \eqref{symp-inst} with $\xi^\Lambda$ given above and verify that this
indeed reproduces \eqref{phiinv}. This provides an independent derivation of this result
using twistor techniques.

The construction of the twistor lines $\txi_\Lambda$ and $\aB$ proceeds by
substituting the transition functions \eqref{symp-inst} (depending on
\eqref{IIBtwi}) into \eqref{eqmuqh}. A straightforward,
although tedious, computation yields the twistor lines in the patch $\cU_{\ob}$
\bea\label{IIBtwist}
\txii{\ob}_a &=&
\frac{\I}{2} \left( \tzeta_a + \frac{\tau_2}{2}
\left( \varpi^{-1} F_a(z)-\varpi \,\bF_a(\bz) \right) \right)
+ \frac{\I}{16 \pi^2} {\sum_{q_a\geq 0}} n_{q_a}^{(0)} q_a \sum_{
n\in \IZ \atop m \not = 0}
\frac{\e^{-S_{m,n,q_a}}}{m  |m \tau + n|} \,  \frac{\varpi^{m,n}_+
+ \varpi}{\varpi^{m,n}_+ - \varpi} \, ,
\nonumber \\
\txii{\ob}_0 &=&
\frac{\I}{2} \left( \tzeta_0 + \frac{\tau_2}{2}
\left( \varpi^{-1} F_0(z)-\varpi \,\bF_0(\bz) \right) \right)
+ \frac{1}{32 \pi^3} {\sum_{q_a\geq 0}} n_{q_a} ^{(0)} \sum_{
n\in \IZ \atop m \not = 0}
\frac{\e^{-S_{m,n,q_a}}}{m  |m \tau + n|^2}
\nonumber
\\
&&\times
\left\{  - \frac{2  \varpi_+^{m,n}\varpi}{(\varpi - \varpi_+^{m,n})^2}
+ \frac{\varpi^{m,n}_+ + \varpi}{\varpi^{m,n}_+ - \varpi}
\[ \frac{m\tau_1 +n}{|m \tau + n|}
+ \I \pi m \tau_2 q_a
\( \frac{z^a}{\varpi^{m,n}_+} + \bz^a \varpi^{m,n}_+ \)
\]
\right\}\, ,
\nonumber
\\
\aB^{[\ob]}  &=&
\frac{1}{4\I}\Bigl[ \sigma + \frac{\tau_2}{2}
\left(\varpi^{-1} W(z)-\varpi \,\bar W(\bz) \right) \Bigr.
\label{IIBtl}
\\
&&\Bigl. \qquad -
\left(\zeta^\Lambda + \frac{\tau_2}{2}
\left( \varpi^{-1} z^{\Lambda} -\varpi \,\bz^{\Lambda}  \right)\right)
\( \tzeta_\Lambda + \frac{\tau_2}{2}
\left( \varpi^{-1} F_\Lambda(z)-\varpi \,\bF_\Lambda(\bz) \right)\)\Bigr]
\nonumber\\
&&
- \frac{1}{32 \pi^3}
{\sum_{q_a\geq 0}} n_{q_a} ^{(0)} \sum_{n\in \IZ \atop m \not = 0}
\frac{\e^{-S_{m,n,q_a}}}{m^2 |m \tau+n|}
\left\{ \frac{\varpi_+^{m,n}+\varpi}{\varpi_+^{m,n}-\varpi }
\right.
\nonumber\\
& &
\left.
-\frac{m\tau_2(m\tau_1+n)}{2|m\tau+n|^2}\(\varpi^{-1}+\varpi\)
-\pi q_a t^a\, \frac{m^2\tau_2^2}{|m\tau+n|}\(\frac{\varpi_+^{m,n}}{\varpi}-\frac{\varpi}{\varpi_+^{m,n}}\)
\right\}\, .
\nonumber
\eea
The result is expressed in terms of Type IIB fields except for
the coordinates $\tzeta_\Lambda$ and $\sigma$ which coincide with
the physical Type IIA fields. This result follows from the gauge equivalence
of the Type IIB twistor space description utilized here with the Type IIA construction
of \cite{Alexandrov:2008gh}, which is demonstrated in Appendices \ref{sect:3} and \ref{sec_Pois}.
It shows that the two twistor spaces are related by a set of gauge transformations
and thus represent two different descriptions of one and the same twistor geometry.

In the presence of instanton corrections, the classical mirror map is insufficient
to relate the remaining Type IIA coordinates $\tzeta_\Lambda$ and $\sigma$
to the Type IIB fields.
The correct relation requires a generalization of \eqref{physmap} and
is provided by the non-perturbative mirror map constructed in the next
section. Together with this map, \eqref{IIBtwist} then provides the twistor lines
underlying the Type IIB hypermultiplet moduli space.

\section{The non-perturbative mirror map}
\label{sec_mirror}

As was shown in the previous section, the mirror map between
the fields entering into the twistor lines $\xi^\Lambda$
does not receive quantum corrections from worldsheet, D($-1)$ and D1-instantons.
On the other hand, the quantum corrections to the twistor lines
$\txi^\Lambda, \aB$ suggest that the map relating the physical
Type IIA fields $\tzeta_\Lambda$, $\sigma$ to $c_\Lambda$, $\psi$ on the Type IIB side will be
subject to similar corrections. The aim of this section is to find the explicit form
of this map. In this course, we follow the same strategy employed
in the construction of the classical mirror map \eqref{physmap},
demanding consistency between the holomorphic action
of SL(2, $\IZ)$ on the complex coordinates $\xi^\Lambda, \txi_\Lambda, \aB$ and the
transformation of the Type IIB twistor lines inherited from the physical Type IIB fields
and the coordinate $\varpi$. As we will see, this condition determines the non-perturbative mirror map
between the physical Type IIA and IIB fields uniquely.

The key ingredient in the construction is the transformation of the Type IIB
fields under SL(2, $\IZ)$. Classically, the transformation properties are inherited
from the dimensional reduction of the ten-dimensional
supergravity action, and we impose that the physical fields parameterizing the quantum corrected
$\cM_K(Y)$ transform according to the classical transformation rules \eqref{SL2Z}. In a sense,
this corresponds to requiring that the SL(2, $\IZ$)-transformations
are realized ``off-shell'', by {\it defining} the physical
fields as the ones which obey these transformation laws
 also at the quantum level.
By demanding that the SL(2, $\IZ$) acts holomorphically on $\xi^\Lambda$, one then establishes that the
SL(2, $\IZ$)-transformation of $\varpi$ is also uncorrected and thus given by \eqref{transz}.

In the next step, we turn to the twistor lines \eqref{IIBtl}
and read $\tilde{\zeta}_\Lambda, \sigma$ as {\it a priori} undetermined
functions of the physical Type IIB fields, which, in the classical
limit, reduce to \eqref{physmap}. Subsequently, we apply the SL(2, $\IZ$) transformations
\eqref{SL2Z} and \eqref{transz} and impose that the transformed
expressions can again be expressed in terms of the holomorphic
twistor lines. This condition suffices
to determine the functions $\tilde{\zeta}_\Lambda, \sigma$ uniquely.
Since their actual derivation is highly technical
and not very illuminating, we will proceed along
another route by first ``guessing" the correct answer and then showing that
the resulting Type IIB twistor lines indeed transform
holomorphically. The generalized mirror map
obtained this way is
\be
\label{instmap}
\begin{split}
\cR& =\frac12\,\tau_2\, ,
\qquad
z^a_{\rm IIA} = z^a_{\rm IIB} \, ,
\qquad
\zeta^0=\tau_1\, ,
\qquad
\zeta^a = - (c^a - \tau_1 b^a)\, ,
\\
\tzeta_a &=  c_a+ \frac{1}{2}\, \kappa_{abc} \,b^b (c^c - \tau_1 b^c)\, + \tzeta_a^{\rm inst},
\qquad
\tzeta_0 =\, c_0-\frac{1}{6}\, \kappa_{abc} \,b^a b^b (c^c-\tau_1 b^c)\, + \tzeta_0^{\rm inst},
\\
\sigma &= -2 (\psi+\frac12  \tau_1 c_0) + c_a (c^a - \tau_1 b^a)
-\frac{1}{6}\,\kappa_{abc} \, b^a c^b (c^c - \tau_1 b^c)\, + \sigma^{\rm inst}\, ,
\end{split}
\ee
with the novel instanton correction terms given by
\bea
\tzeta_a^{\rm inst} &=&
\frac{1}{8\pi^2}
\sum_{q_a \ge 0} n_{q_a}^{(0)}q_a\sum\limits_{n\in \IZ \atop m \not = 0}
\frac{m\tau_1+n}{m|m \tau +n|^2}\,\e^{-S_{m,n,q_a}},
\nonumber
\\ \nn
\tzeta_0^{\rm inst} &=&
-\frac{\I}{16\pi^3}\sum_{q_a \ge 0} n_{q_a}^{(0)}\sum\limits_{n\in \IZ \atop m \not = 0}
\[\frac{(m\tau_1+n)^2}{|m\tau+n|^3}+2\pi q_a\( t^a-\I b^a\,\frac{m\tau_1+n}{|m\tau+n|} \)\]
\frac{\e^{-S_{m,n,q_a}}}{m|m\tau+n|},
\\
\sigma^{\rm inst} &=& \tau_1 \tzeta_0^{\rm inst} - \( c^a - \tau_1 b^a \) \tzeta^{\rm inst}_a
+\frac{\I \tau_2^2}{8\pi^2}\sum_{q_a \ge 0} n_{q_a}^{(0)} q_a t^a  \sum_{n\ne 0}\frac{\e^{-S_{0,n,q_a}}}{n|n|}
\label{mirtzetaz}
\\
&& -\frac{\I}{8\pi^3}\sum_{q_a \ge 0} n_{q_a}^{(0)}
\sum\limits_{n\in \IZ \atop m \not = 0}
\(2-\frac{(m\tau_1+n)^2}{|m\tau+n|^2}\)
\frac{(m\tau_1+n)\e^{-S_{m,n,q_a}}}{m^2|m\tau+n|^2} \, .
\nonumber
\eea
This map constitutes the main result of our paper. Notice that
only the relations between the $\tzeta_\Lambda, \sigma$ and
$c_\Lambda, \psi$ are subject to corrections. In particular
the classical mirror map $z^a_{\rm IIA} = z^a_{\rm IIB}$, which relates
the complex structure moduli to the complexified K\"ahler moduli,
remains uncorrected. Furthermore, since expressing the correction terms
in terms of the Type IIA fields involves the uncorrected relations in \eqref{instmap} only,
the map is easily inverted to give the physical Type IIB fields in terms of
the Type IIA coordinates.  

In order to show that this is indeed the correct relation,
we first substitute \eqref{instmap} into \eqref{IIBtl}.
This gives the final expression for the Type IIB twistor lines:
\bea
\txii{\ob}_a& =&
\frac{\I}{2}\,c_a+ \frac{\I}{4}\, \kappa_{abc} \,b^b (c^c - \tau_1 b^c)
-\frac{\I\tau_2}{8}\, \kappa_{abc}\left( \varpi^{-1} z^b z^c - \varpi \, \bz^b\bz^c \right)
\nn \\
&&
+\frac{\I\tau_2}{16\pi^2}
\sum_{q_a\geq 0} n_{q_a}^{(0)}q_a{\sum_{m,n}}' \,
\frac{1+\varpi_+^{m,n}\varpi}{\varpi-\varpi_+^{m,n}}\,
\frac{\e^{-S_{m,n,q_a}}}{|m\tau+n|^2} ,
\label{newtxinew}
\eea
\bea
\txii{\ob}_0&=&
\frac{\I}{2}\,c_0-\frac{\I}{12}\, \kappa_{abc} \,b^a b^b (c^c-\tau_1 b^c)
+\frac{\I\tau_2}{24}\, \kappa_{abc}\left( \varpi^{-1} z^a z^b z^c - \varpi \,\bz^a \bz^b\bz^c \right)
\nn \\
&&
+\frac{\tau_2}{32\pi^3}
{\sum_{q_a\geq 0}} n_{q_a}^{(0)}{\sum\limits_{m,n}}'
\(\frac{1}{m\xi^0+n}+\frac{m\tau_1+n}{|m\tau+n|^2}\)
\frac{1+\varpi_+^{m,n}\varpi}{\varpi-\varpi_+^{m,n}}\,
\frac{\e^{-S_{m,n,q_a}}}{|m\tau+n|^2}
\nn \\
&&
+\frac{\tau_2}{16\pi^2}\sum_{q_a\geq 0} n_{q_a}^{(0)}q_a{\sum\limits_{m,n}}'
\( t^a\, \frac{1-\varpi_+^{m,n}\varpi}{\varpi-\varpi_+^{m,n}}
-\I b^a \, \frac{1+\varpi_+^{m,n}\varpi}{\varpi-\varpi_+^{m,n}} \)
\frac{\e^{-S_{m,n,q_a}}}{|m\tau+n|^2}.
\label{newtxizeronew}
\\
\aB^{[\ob]} &=&\tfrac{\I}{2}\[\psi+  c_\Lambda \zeta^\Lambda
+ \tfrac{\tau_2}{2} c_\Lambda (\varpi^{-1} z^\Lambda - \varpi \bar{z}^\Lambda ) \]
+\tfrac{\I}{48}\,\tau\tau_2\, \kappa_{abc}\[ \varpi^{-2}z^a z^b z^c +\varpi^2 \bz^a\bz^b\bz^c \]
\nn \\[1.2ex]
&&
+\tfrac{\I}{24}\kappa_{abc} b^a \[\tau_2^2(3t^b t^c+ b^b b^c)- \[2 \zeta^b  + 2 \tau_2 \(\varpi^{-1}-\varpi\) b^b
+ 3 \I \tau_2 \(\varpi^{-1}+\varpi\) t^b \] \zeta^c\]
\nn \\[1.2ex]
&&
+\frac{\tau_2^2}{64 \pi^3} {\sum_{q_a\geq 0}} n_{q_a}^{(0)}
{\sum_{m,n}}'\((m\tau_1+n)\( \varpi^{-1}-\varpi\)-2m\tau_2\)
\frac{1+\varpi_+^{m,n}\varpi}{\varpi-\varpi_+^{m,n}}\,
\frac{\e^{-S_{m,n,q_a}}}{|m\tau+n|^4} . \; \;
\label{twhatalpha}
\eea
Here we abbreviated $\zeta^\Lambda = \( \tau_1 \, , \, - (c^a - \tau_1 b^a) \)$
together with $z^\Lambda = \( 1 , b^a + \I t^a \)$ for convenience.
The SL(2, $\IZ$)-transformations of these twistor lines are
readily obtained by applying \eqref{SL2Z} and \eqref{transz} together with
the intermediate formulas collected in Appendix \ref{ap_transform}.
It turns out, that they transform
according to the {\it classical} law \eqref{SL2Zxi}.
This is highly non-trivial, since the
derivation of the mirror map only imposed that
the twistor lines transform holomorphically under SL(2, $\IZ$)
without specifying the transformation to be of the form \eqref{SL2Zxi}.
This establishes that the holomorphic SL(2, $\IZ$)-action on the twistor space
is realized ``off-shell'' in a sense that it is not modified in the presence
of  worldsheet nor D($-1$) and D1-brane instantons.
Furthermore, this result confirms the correctness of the found mirror map.
The uniqueness of \eqref{instmap} can be established by
adding additional functions of the Type IIB fields to \eqref{mirtzetaz}.
The consistency of the transformations then imposes that these
extra contributions have to vanish identically.

\section{Perturbative singularities and instanton corrections}
\label{sec_reg}

One of the salient features of the non-perturbative instanton
corrections discussed in this paper is their ability to dynamically cure
singularities in the perturbative metric on $\cM$. The prime
example for such a behavior is the conifold singularity
which is smoothed out by D2-brane instantons
wrapping the shrinking cycle \cite{Ooguri:1996me,Saueressig:2007dr}.
Motivated by this observation, we will investigate the interplay between
the D($-1$) and D1-instanton corrections and the
generic singularities of $\cM$ induced by the one-loop correction
\cite{Robles-Llana:2006ez,Alexandrov:2007ec}.
In this course, we will assume that we
work at a regular point in the K\"ahler/Complex structure moduli space, excluding
singularities arising from shrinking (sub-)cycles. Furthermore,
our prime focus will be on the D($-1)$-corrections and we will comment about the
D1-instanton effects only at the end of this section.

\subsection{The perturbative one-loop singularity}

The perturbatively corrected hypermultiplet metric has been given in
\cite{Robles-Llana:2006ez,Alexandrov:2007ec} and its description in terms of
the twistor space was obtained in \cite{Alexandrov:2008nk}.
In terms of the physical Type IIA fields the metric reads
\bea
\de s^2&=&\frac{r+2c}{r^2(r+c)}\,\de r^2
-\frac{1}{r} \(N^{\Lambda\Sigma}-\frac{2(r+c)}{r K} \,z^\Lambda \bz^\Sigma\)
\( \de \tzeta_\Lambda - F_{\Lambda\Theta} \de \zeta^\Theta\)
\(\de \tzeta_\Sigma-\bF_{\Sigma\Xi}\de \zeta^\Xi\)
\nonumber \\
&&
+ \frac{r+c}{16 r^2(r+2c)}\(\de\sigma+\tzeta_\Lambda d \zeta^\Lambda
- \zeta^\Lambda \de \tzeta_\Lambda+4c\, \cA_K \)^2
+\frac{4(r+c)}{r}\,\cK_{a{\bar b}}\,\de z^a \de \bz^{\bar b} \, ,
\label{hypmet}
\eea
where $r=\e^\phi$, $c = -\frac{\chi_X}{192 \pi}$,
$N_{\Lambda\Sigma} \equiv \I (F_{\Lambda\Sigma} - \bar F_{\Lambda\Sigma})$,
$\cK=-\log K(z,\bz)$ is the K\"ahler potential of the special \kahler base $\cK_K(X)$
and $\cA_K \equiv \I\(\cK_a \de z^a -\cK_{\bar a} \de \bz^{\bar a}\)$
is its K\"ahler connection.

With respect to the string coupling $r$, the metric possesses three
apparent singularities at $r=0$, $r=-c$ and $r=-2c$. The last two
arise from the one-loop correction and occur in CY compactifications
with positive Euler number $\chi_X > 0$.
Notably, the first two points constitute coordinate singularities only. The singularity at
$r=0$ can be removed by a simple rescaling of $\zeta^\Lambda,\ \tzeta_\Lambda$ and $\sigma$
by a power of $r$ and returning to the variable $\phi$, whereas the singularity at $r=-c$
disappears after one trades $r$ for the ten-dimensional string coupling (cf. \eqref{phipertB})
\be
\tau_2
= 4 \,e^{\frac12 \, \cK(z,\bar z)} \, \sqrt{r+ c} \, .
\label{taur}
\ee
This picture is confirmed by computing the quadratic curvature invariant
$R_{\mu\nu\rho\sigma} R^{\mu\nu\rho\sigma}$, which diverges at $r = -2c$ only,
and remains regular at $r=-c$ and $r=0$. Thus the only curvature
singularity of the perturbatively corrected hypermultiplet
metric appears for $\chi_X > 0$ at $r=-2c$.\footnote{This observation is already
suggested by Fig. 1 of \cite{Davidse:2005ef}.}

At the level of the twistor space construction this singularity is
caused by the degeneration of the basis of holomorphic $(1,0)$-forms
\eqref{defPi}. Evaluating these for the
perturbative twistor lines \eqref{gentwi} and taking suitable linear
combinations, an explicit basis of $(1,0)$-forms is given by
\cite{Alexandrov:2008nk}
\be
\begin{split}
&\Pi^a = \de z^a,
\qquad
\cY_\Lambda =
\de\tzeta_\Lambda-F_{\Lambda\Sigma}\de\zeta^\Sigma,
\\
\Sigma= \de \e^{\phi} &+2 c \, \de \log\tau_2
+\frac{\I}{4}\[\de \sigma+\tzeta_\Lambda\de \zeta^\Lambda-\zeta^\Lambda\de \tzeta_\Lambda\].
\end{split}
\label{oneformscalss}
\ee
The dilaton-direction in \eqref{hypmet} is generated by the real part of $\Sigma$.
This part degenerates at the point where
\be\label{cond2}
\de \e^{\phi} +2c \, \de \log\tau_2 = 0 \qquad {\mbox{mod}} \; \; \Pi^a, \cY_\Lambda \, .
\ee
Substituting the relation \eqref{taur},
one immediately finds that the only solution of \eqref{cond2} is given by $r=-2c$, which
clarifies the origin of the perturbative singularity from the twistorial viewpoint.

\subsection{The effect of D-instanton corrections}

In order to analyze the effects of the D-instanton contributions
on the perturbative singularity, we first
compute the D($-1$) and D1-instanton corrections to \eqref{oneformscalss}. Starting from the Type IIA twistor
lines \eqref{xiqlineB} and using the transition functions \eqref{gensymp}
together with \eqref{defrho} allows to determine the non-perturbative corrections
to $\alpha_0^{[+]}, \txi^{[+]}_{\Lambda,0}$. Substituting
the result into \eqref{defPi}, the instanton-corrected basis of
holomorphic $(1,0)$-forms is given by
\bea \nn \label{holinst}
 \Pi^a & = & \de z^a \, , \qquad
\cY_\Lambda  =
\de\tzeta_\Lambda-F_{\Lambda\Sigma}\de\zeta^\Sigma
-\frac{\I}{4\pi^2}\sum\limits_{\gamma} \hng{} \,q_{\Lambda}\,\de \Kkl \, ,
\nn \\
 \Sigma & = & \de \e^{\phi} +2c \, \de \log\tau_2 \\ \nn
&& +\frac{\I}{4}\Big[\de \sigma+\tzeta_\Lambda\de \zeta^\Lambda-\zeta^\Lambda\de \tzeta_\Lambda
-\tfrac{\I}{8\pi^2}\sum\limits_{\gamma} \hng{}
q_\Lambda\(\tau_2 z^\Lambda \de\cLp-\cLm \de\(\tau_2\bz^\Lambda\) \) \Big].
\nn
\eea
Here
\be
\cLp\equiv \left. \frac{\I}{4}\,\frac{\de}{\de\varpi}\Ikl^{(1)} \right|_{\varpi = 0}
\ee
denotes the subleading coefficient in the $\varpi$-expansion of $\Ikl^{(1)}(\varpi)$
around $\varpi = 0$,
$\Kkl$ and $\Ikl^{(1)}$ are defined in \eqref{IKkl}, and $\e^{\phi}$
is the instanton corrected contact potential
\eqref{phiinstfull}. Owed to the relation $q_\Lambda z^\Lambda \cLp=q_\Lambda \bz^\Lambda \cLm$,
the term appearing in the square bracket in $\Sigma$ is real. Thus the equation controlling
the degeneracy of the basis \eqref{holinst} is still of the form \eqref{cond2}, with the perturbative $\e^{\phi}$
now dressed up with instanton corrections.

In order to understand the fate of the perturbative singularity at $r = -2c$, we
need to understand the behavior of $\e^{\phi}$ at strong coupling $\tau_2 \rightarrow 0$.
At this point it is useful to switch to the mirror symmetric Type IIB description
\eqref{phiinv} where we can use $S$-duality to relate $\e^{\phi}$ at strong and weak
string coupling. The D($-1$)-instanton corrections (leaving out the D1-instanton contribution for the
time being) to the four-dimensional dilaton are readily obtained
from the second term in \eqref{phiinv} by setting
$q_a = 0$, $n_0^{(0)} = -\chi_Y/2$. Together with the perturbative worldsheet
and string loop corrections, they combine into a real analytic Eisenstein series
\be
\e^{\phi_{\rm D(-1)}} = -\frac{\chi_Y \tau_2^{1/2}}{16(2\pi)^3}\, \cE_{3/2}(\tau,\bar\tau) \, ,
\qquad \cE_{3/2}(\tau,\bar\tau) := {\sum\limits_{m,n}}' \, \frac{\tau_2^{3/2}}{|m \tau + n|^3} \, .
\label{contrD}
\ee
The strong coupling limit of these terms can then be extracted by using the SL(2, $\IZ$)-invariance
of $\cE_{3/2}$ and can conveniently be done by first applying an $S$-duality $\tau \mapsto - \tau^{-1}$
to the weak coupling expansion of $\cE_{3/2}$ and subsequently taking the limit $\tau_1 \rightarrow 0$
\be
\e^{\phi_{\rm D(-1)}} = -\frac{\chi_Y}{16(2\pi)^3}\, \( 2 \zeta(3) \tau_2^{-1} + 4 \zeta(2) \tau_2 \)
+ O(\e^{-2 \pi \tau_2^{-1}}) \, .
\ee
Including the
tree-level term, $\e^{\phi}$ then has the strong coupling expansion
\be\label{phi:strong}
\e^{ \phi} = -\frac{\chi_Y}{16(2\pi)^3}\, \( 2 \zeta(3) \tau_2^{-1}
+ 4 \zeta(2) \tau_2\) + \hf\, V(t)\tau_2^2 + \ldots\, ,
\ee
where the dots stand for terms which are exponentially suppressed. As expected,
the strong coupling asymptotics is dominated by the
instanton effects. In this asymptotics 
the four-dimensional dilaton (related to the four-dimensional
string coupling via $\e^{-\phi} \propto g_{(4)}^2$) behaves
as $\e^\phi\sim \tau_2^{-1}$ so that $g_{(4)}\sim g_{(10)}^{-1/2}$.
In other words, the four-dimensional coupling is prohibited from divergence:
the region of large $g_{(4)}$ is inaccessible on the studied corners of the moduli space.

In the Type IIB description, the perturbative singularity, $r=-2c$ with $c = \frac{\chi_Y}{196 \pi}$, appears
for CYs $Y$ (mirror to $X$) with $\chi_Y < 0$. In order to make a statement about the fate of this singularity in
the presence of D($-1)$-instantons, the condition \eqref{cond2} indicates that it is sufficient to
consider the asymptotics of  $\frac{\de \e^\phi}{\de \log\tau_2}$
in the weak and strong coupling regime.
For $\tau_2$ large (and in the large volume limit), the contact potential is dominated
by the classical term so that
\be
\frac{\de \e^\phi}{\de \log\tau_2}\ \mathop{\sim}\limits_{\tau_2\to\infty}\ V(t)\, \tau_2^2 > 0.
\ee
At strong coupling, the expansion \eqref{phi:strong} yields
\be
\frac{\de \e^\phi}{\de \log\tau_2}\ \mathop{\sim}\limits_{\tau_2\to 0}\
\frac{\zeta(3)\chi_Y}{8(2\pi)^3}\,  \tau_2^{-1}.
\ee
For $\chi_Y < 0$ the two asymptotics have opposite signs. Therefore
the equation \eqref{cond2} necessarily has a solution at finite value of $\tau_2$.
Thus we conclude that the D($-1$)-instantons {\it do not resolve} the singularity of perturbative
hypermultiplet metric.

Before closing this section, let us briefly comment on the effect of the D1-instantons at strong
coupling. Their contributions are given by the terms with non-vanishing charge $q_a$ in \eqref{phiinv}.
In this case, the application of $S$-duality does not lead to terms which are exponentially
suppressed as $\tau_2 \rightarrow 0$, since the SL(2, $\IZ$)-transformations also act on the
other fields, in particular $t^a \mapsto |c \tau + d| t^a$.
Therefore, one should work directly with the double
sums in \eqref{phiinv}.
The leading contribution at small $\tau_2$ (again assuming
$\tau_1=0$) comes from the terms with $n=0$ and is given by
\be\label{breaksdown}
\frac{\tau_2^{-1}}{4(2\pi)^3}\sum_{\hat\gamma_+} n_{q_a}^{(0)}
\sum_{m=1}^{\infty} \frac{e^{-2\pi m\tau_2 q_a t^a}}{m^3}
\sim \frac{\zeta(3)}{4(2\pi)^3}\,\tau_2^{-1} \sum_{\hat\gamma_+}n_{q_a}^{(0)}.
\ee
In order for the sum over the charges $q_a$ to converge, the limit $\tau_2 \rightarrow 0$ 
has to be taken by keeping $q_a t^a \tau_2$ fixed and sufficiently large.
In this ``decompactification limit'' one concludes that the D1-instanton 
contribution to the contact potential
has the same asymptotics as the one due to D($-1$)-instantons. 
Keeping $t^a$ finite, however,
the sum over BPS invariants $n_{q_a}^{(0)}$ diverges, 
so that it is hard to draw any definite conclusions.
A proper treatment of this limit
will, most likely, involve a resummation of the instanton
series, as, e.g., along the lines suggested in \cite{Pioline:2009ia}.
However, in this work we are not embarking on this point.

\section{Discussion and outlook}
\label{sec:7}

The main result of our paper is the non-perturbative 
mirror map \eqref{instmap} which  establishes a relation between
Type IIA and Type IIB string theory
compactified on a generic pair of mirror Calabi-Yau threefolds,
taking into account worldsheet, D($-1$), and D1-instanton
corrections. This map
constitutes a non-perturbative generalization of the
classical limit obtained in \cite{Bohm:1999uk}.
In contrast to the classical case, our result is derived
from the twistor space description
of the corresponding hypermultiplet moduli spaces, thereby
avoiding the explicit construction
of the underlying QK metrics.  Notably, the quantum corrections
to the classical map are uniquely determined
by the consistent implementation of symplectic covariance (Type IIA) and
SL(2, $\IZ$)-transformations (Type IIB) on the twistor space.
As a spin-off we found
that the SL(2, $\IZ$)-transformation of the twistor lines
does not receive quantum corrections from these non-perturbative effects.
We expect that this result will continue to hold once
the additional corrections from D3, D5, and NS5-instantons
are included. The ``off-shell" realization of the SL(2, $\IZ$)
invariance could then provide a powerful tool in
unraveling the physical structures underlying these corrections.

A natural question arising from our result
concerns the inclusion of these additional corrections
in the generalized mirror map. Using the Type IIA formulation,
the twistor lines describing  D2-branes wrapped on arbitrary
three-dimensional special Lagrangian submanifolds have
been calculated explicitly in the linear instanton
approximation \cite{Alexandrov:2008gh} and
to all orders in a somewhat implicit form in \cite{Alexandrov:2009zh}.
The resummation technique of Appendix \ref{sec_Pois} can, in principle,
be applied to this case as well, thereby providing an
interesting generalization of the results reported here.
An important test for the consistency
of the resulting Type IIB twistor lines is then given by
their transformation under SL(2, $\IZ$).
In particular the D3-instantons mirror to the B-type D2-instantons
are expected to organize themselves into a modular form. This would
allow the generalization of our construction,
taking these additional corrections into account as well.

Curiously, a ``naive" Poisson resummation of the Type IIA twistor lines in the presence
of B-type D2-instantons does not lead to Type IIB twistor lines exhibiting the desired
behavior under SL(2, $\IZ$). One possible explanation
for this intriguing observation is that the Type IIA instanton numbers $n_\gamma$ develop
a dependence on the resummed charge once all D2-instantons are included.
Thus further progress in this direction should go hand in hand with a better
understanding of the instanton numbers appearing on the Type IIA side together
with their mirrors.

Our second result concerns the singularity structure of the hypermultiplet
moduli space. In this context, we found that the singularities
in the hypermultiplet moduli space occurring at the perturbative
level are not resolved by the inclusion of D($-1$)-brane instantons,
even though they are part of the same modular invariant. This is,
however, in good agreement with the expectation that the dominating non-perturbative
contribution at strong coupling should be given by the NS5-brane instantons which
have not been included in our analysis. Some progress towards understanding the role
of the NS5-brane instantons has been made in \cite{Alexandrov:2006hx,Pioline:2009qt},
but their contribution remains to be fully understood. We hope to return to this point in the future.

\section*{Acknowledgements}

The authors are grateful to Boris Pioline and Stefan Vandoren for very useful discussions.
The research of S.A. is supported by CNRS.
F.S.\ acknowledges financial support from the ANR grant BLAN06-3-137168
and thanks the LPTA at Montpellier for hospitality while the work was completed.

\appendix

\section{Type IIA and Type IIB twistor spaces and their relation}
\label{sect:3}

Keeping key features as, e.g., the SL(2, $\IZ$)-invariance
on the Type IIB side or symplectic invariance and wall-crossings
in Type IIA manifest, naturally leads to twistor space descriptions
of instanton corrected HM moduli spaces, which utilize different sets of patches and transition functions.
Indeed, the Type IIB description presented in Section \ref{sec_instA} is quite different
from its Type IIA cousin constructed in \cite{Alexandrov:2008gh}
and reviewed in Appendix \ref{subsec_linins}.
However, the mirror symmetry indicates that there should be an intrinsic relation
between these constructions.
In Appendix \ref{ssect:gt}, we will then show that the two
constructions are indeed equivalent and
related by a gauge-transformation.
Technical details of the calculation are further referred to Appendix \ref{sec_Pois}.

\subsection{Instanton corrected Type IIA HM moduli space}
\label{subsec_linins}

The HM moduli space of Type IIA strings compactified on a CY $X$ receives instanton
corrections from D2-branes wrapping the 3-cycles of $X$. The subclass of these D2-instanton
corrections wrapping A-cycles is
related to D($-1$) and D1-instantons by mirror symmetry \cite{Ooguri:1996ck}.
In particular, this implies that the contact potential underlying the Type IIA picture
can be obtained by Poisson resumming \eqref{phiinv}
on $n \in \IZ$ and subsequently applying the classical mirror map \eqref{physmap},
which, for the fields appearing in \eqref{phiinv}, does not receive quantum
corrections. As a result, one obtains \cite{RoblesLlana:2007ae}
\be
\begin{split}
e^{\Phi_{\rm IIA}} =& \frac{\tau_2^2}{16}\, K(z,\bz)+\frac{\chi_X}{192\pi}
\\
& +\frac{\tau_2}{16\pi^2}{\sum\limits_{\gamma}} \hnkl\sum\limits_{m> 0}
\frac{|q_\Lambda z^\Lambda|}{m}\, \cos\(2\pi m q_\Lambda \zeta^\Lambda\)
K_1(2\pi m \tau_2 |q_\Lambda z^\Lambda|)\, .
\end{split}
\label{phiinstfull}
\ee
Matching \eqref{phiinv} and  \eqref{phiinstfull}
 requires the summation over the charge lattice
$\gamma=\{q_0,q_a\}$ where $q_0\in \IZ, \ q_a\gamma^a \in H_2^+(Y) \cup H_2^-(Y) \cup \{0\}$
excluding the case $\gamma=0$,
and implies that the instanton numbers $n_\gamma$ are related
to the genus zero Gopakumar-Vafa invariants {\it of the mirror CY} by
\be\label{instnr}
n_\gamma = \hn_{(q_0,\pm q_a)} \equiv n_{q_a}^{(0)}(Y) \quad\mbox{\rm for}\quad \{ q_a \} \ne 0\, , \qquad
\hn_{(q_0,0)}=2n_{0}^{(0)}= \chi_X\, .
\ee

\begin{figure}[t]
\renewcommand{\baselinestretch}{1}
\begin{center}
\leavevmode
\put(-215,6){\includegraphics[width=0.40\textwidth]{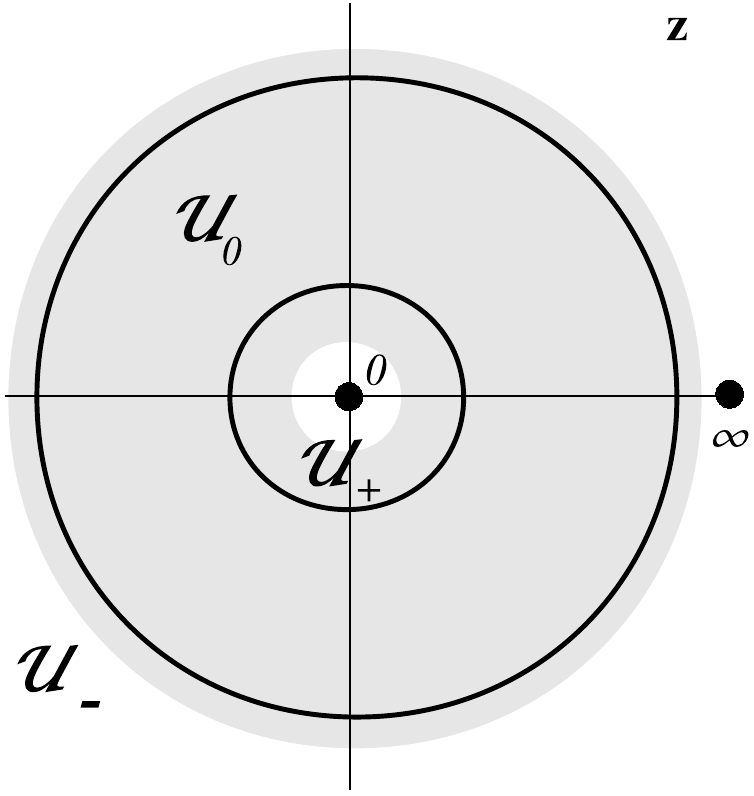}}
\leavevmode
\put(-30,95){\includegraphics[width=0.10\textwidth]{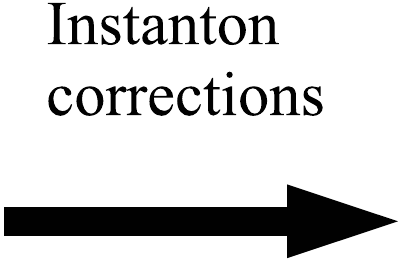}}
\leavevmode
\put(30,4){\includegraphics[width=0.40\textwidth]{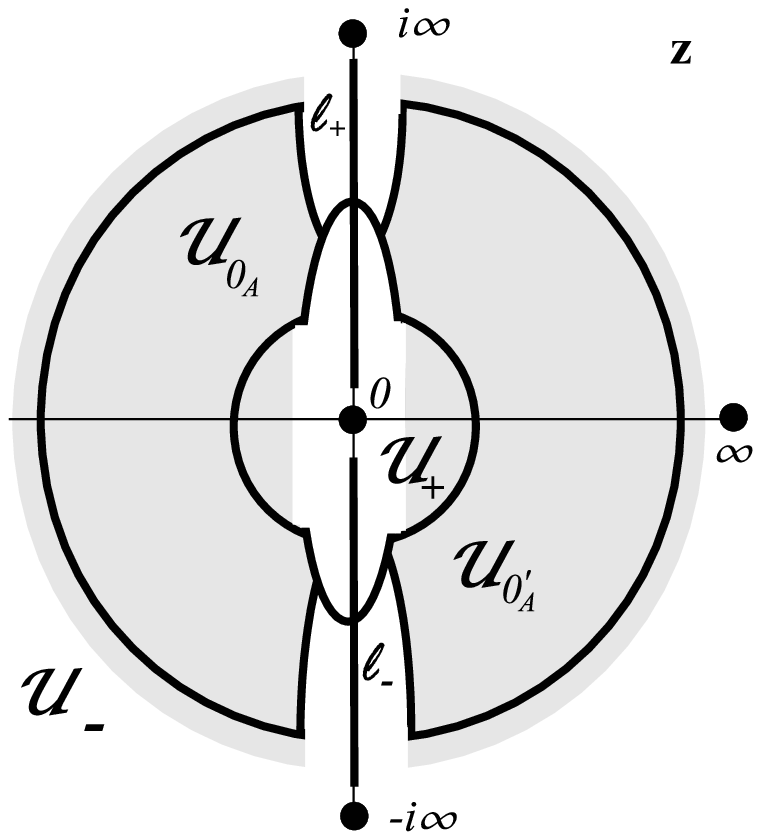}}
\end{center}
\parbox[c]{\textwidth}{\caption{\label{Fig.2}{Transition from the classical to
the instanton corrected Type IIA twistor space constructed in \cite{Alexandrov:2008gh}.
The instanton corrections are encoded in the two BPS rays
$\ell_{\pm}$ which are covered by an extension of the patches $\cU_{\pm}$. The description
is gauge-equivalent to the novel Type IIB twistor space illustrated in Fig.\ 1.}}}
\end{figure}

The twistor space description for these corrections has recently been
developed in \cite{Alexandrov:2008gh,Alexandrov:2009zh}
and the covering underlying the construction is shown in Fig.\ \ref{Fig.2}.
It consists of the usual patches around the poles $\cU_\pm$ and two additional patches,
$\cU_{\oa}$ and $\cU_{\opa}$, which cover the left and the right half-planes of $\CP$ considered
as a complex $\varpi$-plane.
They are separated by
two rays joining $\varpi=0$ and $\varpi=\infty$ and going along
the semi-infinite imaginary axes $\ell_\pm\equiv  \I\IR^{\pm}$.

The discontinuities along $\ell_{\pm}$ imply the existence of
two transition functions relating $\cU_{\oa}$ and $\cU_{\opa}$
since the analytical continuation from one patch to another
can be done either through $\ell_+$ or $\ell_-$.
These two functions read, respectively, as
\be
\label{symp2-inst}
H^{[00']_+}=\frac{\I}{2}\,G_{\rm IIA}(\xi)   \, , \qquad
H^{[00']_-}=-\frac{\I}{2}\,\bG_{\rm IIA}(\xi) \, ,
\ee
where the function
\be
G_{\rm IIA}(\xi)=
\frac{1}{(2\pi)^2}
\ {\sum\limits_{\scriptsize\lefteqn{\gamma_+}}}\,\hnkl\,
\Li_2\left(e^{-2\pi \I q_\Lambda \xi^\Lambda } \right)\, 
\label{prepH}
\ee
incorporates the instanton contributions.
Here the sum over $\gamma_+$ is supported on charges $\gamma=\{q_\Lambda\}$
with $\Re \left( q_\Lambda z^\Lambda \right)>0$.
In addition one has to specify the transition functions connecting $\cU_{\oa}$ (or $\cU_{\opa}$)
to $\cU_{\pm}$. They have both perturbative and instanton contributions and read
\be
\label{gensymp}
\begin{split}
\hHij{+0}&
=  \frac{\I}{2}\( F(\xi)
+\cG(\xi)
\) ,
\qquad
\hHij{-0}
=   \frac{\I}{2}\( \bF(\xi)
-\cG(\xi)
\),
\end{split}
\ee
where we introduced
\be
\cG(\xi)=\frac{\I}{4\pi^3}\ {\sum\limits_{\scriptsize\lefteqn{\gamma_+}}}\,\hnkl\,
\int_0^{-\I\infty} \frac{\Xi\,\de\Xi}{(k_\Lambda \xi^\Lambda)^2-\Xi^2}\, \Li_2\left(e^{-2\pi \I\, \Xi} \right)\, .
\label{funGg}
\ee
Note that the function \eqref{prepH} and its conjugate are simply
the discontinuities of $\Gg$ along the cuts which near the
poles can be taken to be along the contours $\ell_\pm$. This property
ensures the mutual consistency of the transition functions introduced above and
allows to reduce all calculations to the evaluation of integrals along ``open contours'' $\ell_\pm$
\cite{Alexandrov:2008gh,Alexandrov:2009zh}.
Finally, as in \eqref{symp-cmap},
there is one non-vanishing anomalous dimension which is $\ci{+}_\alpha= \chi_X/(96\pi)$.
It incorporates the effect of the one-loop perturbative contribution
found in \cite{Robles-Llana:2006ez}.
Altogether one can check that the presented construction reproduces
the contact potential \eqref{phiinstfull}.

The twistor lines in the patches $\cU_{\oa}$ and $\cU_{\opa}$
have been computed in \cite{Alexandrov:2008gh}
and are given by
\bse
\label{xiqlineB}
\bea
\label{xiqlineB2}
\xi^\Lambda &=& \zeta^\Lambda + \cR \left(
\varpi^{-1} z^\Lambda - \varpi \, \bz^\Lambda\right) 
\\
\label{txiqlineB2}
\tilde{\xi}^{[0_A]}_\Lambda &=&
\frac{\I}{2}\(\tzeta_\Lambda
+\cR \left( \varpi^{-1} F_\Lambda - \varpi \, \bF_\Lambda \right) \)
+\frac{\I}{32\pi^2}{\sum\limits_{\gamma}} \,\hnkl\, q_\Lambda\, \Ikl^{(1)}(\varpi)\, ,
\\
\label{txifqlineB2}
\aA^{[0_A]} &=& \sigma
+\cR (\varpi^{-1} W-\varpi \,\bar W) +\frac{\I\chi_X}{24\pi} \log \varpi
+\frac{\I\cR}{2\pi^2}{\sum\limits_{\gamma}}  \, \hnkl q_\Lambda
\(\varpi^{-1}z^\Lambda+\varpi\bz^\Lambda \)\Kkl
\nn\\
&& + \frac{1}{16\pi^2}
{\sum\limits_{\gamma}}  \hnkl\[\frac{1}{\pi \I}\,
\Ikl^{(2)}(\varpi)+q_\Lambda \xi^\Lambda \Ikl^{(1)}(\varpi) \]\, ,
\eea
\ese
where the sum over $\gamma$ runs over the union
of $\gamma_+$ and $\gamma_-$ and we introduced
\be
\begin{split}
\Kkl\equiv &
\frac{\I}{4}\,\Ikl^{(1)}(0) =
\sum\limits_{m=1}^{\infty} \frac{1}{m} \sin\(2\pi m q_\Lambda \zeta^\Lambda\)\,
K_0\(4\pi m \cR|q_\Lambda z^\Lambda|\)\, ,
\\
\Ikl^{(\nu)}(\varpi)
\equiv &
\sum_{m=1}^{\infty} \sum_{s=\pm 1} \frac{s^\nu}{m^\nu}\,
e^{-2\pi \I s m q_\Lambda \zeta^\Lambda }
\int_{0}^{\infty}\frac{\d t}{t}\, \frac{t-\epskl s\I \varpi}{t+\epskl s\I\varpi}\,
e^{-2\pi m\epskl\cR q_\Lambda \( t^{-1} z^\Lambda +t \bz^\Lambda\)} \, ,
\end{split}
\label{IKkl}
\ee
with $\epskl =\sign (\Re q_\Lambda z^\Lambda)$.

We remark that the only effect of the non-vanishing anomalous dimension
on the twistor lines is the logarithmic term in $\aA$ which is present in all patches.
If the cut of the logarithm is set to be along the imaginary axis, one gets an additional contribution to
the transition function $\Hij{00'}$,
which is needed to cancel the difference between the two branches of the logarithm on the two sides of the cut.
To relate our picture to the Type IIB formulation in
the next subsection, it will be convenient to split the logarithmic term into two parts
and direct the cut of each term along positive and negative imaginary half-axes, respectively.
In this case, the additional contributions read\footnote{As a consequence of this anomalous contribution,
the consistency condition \eqref{consisth} picks up an additional
constant term when relating patches separated by the logarithmic branch-cut.}
\be
H^{[00']_\pm}_{\rm an}=\pm\frac{\I\chi_X}{96}\, .
\label{anomal00}
\ee
This anomalous contribution is important for
establishing mirror symmetry at the level of the twistor space.

\subsection{Mirror symmetry of the twistor spaces}
\label{ssect:gt}

We will now show that the two twistor spaces underlying the Type IIA
and Type IIB formulations are actually identical, as required by mirror
symmetry. More precisely, we demonstrate that they are related
by a gauge transformation of the form \eqref{gaugeeq}. Since the twistor lines $\xi^\Lambda$
are already identical in the Type IIA and Type IIB formulation, it suffices to consider gauge transformations
of $\txi^{[i]}_\Lambda$ and $\aB^{[i]}$
\be\label{twigauge}
\txi^{[i]}_\Lambda \mapsto \txi^{[i]}_\Lambda - \p_{\xi^\Lambda} T^{[i]}\, ,
\qquad
\aB^{[i]} \mapsto \aB^{[i]} - T^{[i]} \, .
\ee
Thus, knowing the relation between the twistor lines $\aB$ in the two constructions
allows us to directly read off the holomorphic functions underlying the gauge transformation.

In the first step towards establishing this gauge equivalence,
we perform a Poisson resummation of the Type IIA twistor lines defined in the patches
$\cU_{\oa}$ and $\cU_{\opa}$. The details of this resummation can be found in Appendix \ref{sec_Pois}.
In both patches the result can be written as
\be\label{IIAtw}
\txii{\oa}_\Lambda = \txii{\ob}_{\Lambda} + \p_{\xi^\Lambda} \cT_\alpha \, ,
\qquad
\aB^{[\oa]}  = \aB^{[\ob]} + \cT_\alpha \, ,
\ee
where $\cT_\alpha(\xi^\Lambda)$ is given in \eqref{eq:master}.
Here the relation is written on the intersection $\cU_{\oa}\cap\cU_{\ob}$ which coincides
with the second quadrant of the complex $\varpi$-plane. To write it in other quadrants,
it is sufficient to put primes on the patch labels in appropriate places.

This relation between twistor lines has the form
of the gauge transformation \eqref{twigauge} so that it is tempting to immediately apply
such interpretation to it. However, this cannot be done straightforwardly because the
function $\cT_\alpha$ is not holomorphic everywhere but has discontinuities
along the real and imaginary axes, originating from the sign functions in \eqref{eq:master}.
This is consistent with the fact that $\cT_\alpha$
arises from the pole at $t=\pm\I\varpi$ in the original integral \eqref{IKkl},
which is responsible for the discontinuity of the Type IIA twistor lines
across the rays $\ell_\pm$ along the imaginary axis.
Similarly, the discontinuity along the real axis accounts for the 
discontinuity of the Type IIB twistor lines owed to the
condensation of poles, as can be seen explicitly from the dual
representation of the $\cT$-terms in \eqref{eq:A21}.

The presence of these discontinuities suggests that one should refine the covering of the $\CP$
used in the Type IIA picture by representing $\cU_{0_A}$ and $\cU_{0'_A}$ as a union of three patches
\be
\cU_{0_A}=\cU_{\bp} \cup \cU_{\rm II}\cup \cU_{\rm III},
\qquad
\cU_{0'_A}=\cU_{\bm} \cup \cU_{\rm I}\cup \cU_{\rm IV},
\ee
which are related by trivial transition functions. Here $\cU_{\bmp}$
surround the positive and negative real axes and
the other patches cover the corresponding quadrants of the $\varpi$-complex plane.
In each quadrant one can define the holomorphic function
$\cT_\alpha^{[i]}=\cT_\alpha$, $i \in \{{\rm I,II,III,IV}\}$,
which then can be analytically continued to the whole plane.
It is easy to check that they are given by
\be
\cT_\alpha^{[{\rm I}]}=\frac{\I}{2}\,G_{\rm A}^-,
\qquad
\cT_\alpha^{[{\rm II}]}=-\frac{\I}{2}\,G_{\rm A}^+,
\qquad
\cT_\alpha^{[{\rm III}]}=\frac{\I}{2}\,\bG_{\rm A}^-,
\qquad
\cT_\alpha^{[{\rm IV}]}=-\frac{\I}{2}\,\bG_{\rm A}^+,
\label{gaugefunq}
\ee
where we defined
\be
G_{\rm A}^\pm(\xi^\Lambda)=
\frac{1}{(2\pi)^2}
\ {\sum\limits_{\scriptsize\lefteqn{_\pm\gamma_+}}}\, n_{q_a} ^{(0)}\,
\Li_2\left(e^{-2\pi \I q_\Lambda \xi^\Lambda } \right)
+\frac{\chi_X}{96}\, ,
\label{prepHpm}
\ee
with the sum over charges spanning the following lattice
\be
_\pm\gamma_+=\{\gamma: \ \Re (q_\Lambda z^\Lambda )>0 \ {\rm and} \ \pm q_a t^a\ge 0\}.
\ee
This definition implies $G_{\rm A}^+  + G_{\rm A}^-=G_{\rm IIA}+\frac{\chi_X}{48}$,
 so that \eqref{gaugefunq} correctly reproduce the discontinuity across the BPS rays $\ell_\pm$.

As a consequence, the relation between twistor lines \eqref{IIAtw} in the four quadrants of
the $\varpi$-plane can be rewritten as
\be\label{IIABtw}
\txi^{[i]_A}_\Lambda = \txi^{[i]_B}_{\Lambda} + \p_{\xi^\Lambda} \cT_\alpha^{[i]} \, ,
\qquad
\aB^{[i]_A}  = \aB^{[i]_B} + \cT_\alpha^{[i]} \, ,
\ee
where the indices $[i]_A$ and $[i]_B$ denote the restriction
of the corresponding Type IIA or Type IIB twistor line to the patch $\cU_i$.
Since every $\cT_\alpha^{[i]}$ is holomorphic in $\cU_i$, they now can be interpreted as
gauge-transformations $T^{[i]}$ from \eqref{twigauge} relating
$\txi^{[i]_A}_\Lambda,\aB^{[i]_A}$ to $\txi^{[i]_B}_\Lambda,\aB^{[i]_B}$.
In this way we recover the Type IIB twistor lines in each of the quadrants.

Such gauge transformations affect essentially all transition functions.
Let us first consider the corresponding change
of the transition functions between quadrants I and II and
between III and IV.
To get the complete result, one should also take into account the effect of the anomalous dimension
expressed in terms of the additional contribution to $\hHij{00'}$ \eqref{anomal00}.
Then from \eqref{symp2-inst}, \eqref{gaugefunq} and \eqref{anomal00}, the new functions are
found to be
\be
\begin{split}
{H}^{[{\rm I\, II}]_B} = -\(H^{[00']_+}+H^{[00']_+}_{\rm an}\)
+ \cT^{[{\rm I}]}_\alpha - \cT^{[{\rm II}]}_\alpha = 0
\, ,
\\
{H}^{[{\rm III\, IV}]_B} = \(H^{[00']_-}+H^{[00']_-}_{\rm an}\)
+ \cT^{[{\rm III}]}_\alpha - \cT^{[{\rm IV}]}_\alpha = 0
\, .
\end{split}
\ee
Thus the gauge transformation removes the branch cuts along the BPS rays.
Therefore, $\cU_{\rm I}$ and $\cU_{\rm II}$, as well as $\cU_{\rm III}$ and $\cU_{\rm IV}$,
can be unified in a single patch, which coincides with $\cU_{\ob}$ and $\cU_{\opb}$, respectively.

Next, the transition functions between the quadrants and
the patches $\cU_{\bmp}$ become
\be
H^{[i \bm]_B}=\cT^{[i]}_\alpha, \quad i={\rm I,IV},
\qquad
H^{[i \bp]_B}=\cT^{[i]}_\alpha, \quad i={\rm II,III}.
\ee
Comparing the representation \eqref{eq:A21} of the function $\cT_\alpha$ with
the function $G_{\rm IIB}$ given in \eqref{prepotGG}, one observes that they are
{\it almost} the same. In this ``Type IIB'' representation, the difference is due
to some sign factors and an additional non-holomorphic term in the exponential of $\cT_\alpha$.
However, performing the Poisson resummation, one finds that the difference between these
two functions is in fact holomorphic in both patches $\cU_{\bmp}$. Indeed, the
resumed expression for $\cT_\alpha$ is given in \eqref{eq:master}, whereas the result
of resummation of $G_{\rm IIB}$ immediately follows from the first equation in \eqref{res:int1}
and reads
\be
G_{\rm IIB}=
\frac{\tve}{4 \pi^2} \sum\limits_{q_a\ge 0  }
n_{q_a}^{(0)}\( \sum\limits_{\tve q_0=1}^\infty \Li_2(\e^{-2 \pi \I  q_\Lambda \xi^\Lambda})
+ \frac{1}{2}\,  n_{q_a}^{(0)}\Li_2\(\e^{-2 \pi \I  q_a \xi^a}\)\) \, .
\label{resGIIB}
\ee
Then, denoting $x=q_a b^a$ and $y=q_a t^a$, for both $i=\,$II and III one obtains
\be
\begin{split}
\cT^{[i]}_\alpha+\frac{\I}{2}\,G_{\rm IIB}= &
\frac{\I}{8\pi^2}\(
\sum_{\hat\gamma:\ x<0,y>0} \sum_{0<q_0<x}
-\sum_{\hat\gamma:\ x>0,y>0} \sum_{-x<q_0<0}\)n_{q_a}^{(0)} \Li_2\(\e^{-2\pi \I  q_\Lambda \xi^\Lambda}\)
\\
&
+\frac{\I}{16\pi^2}\(
\sum_{\hat\gamma:\ x<0,y>0}
-\sum_{\hat\gamma:\ x>0,y>0} \)n_{q_a}^{(0)}\Li_2\( e^{-2\pi \I q_a\xi^a}\).
\end{split}
\label{diffTG}
\ee
This is a holomorphic function and therefore can be removed by a gauge transformation
in the patch $\cU_{\bp}$. Similarly, one can show that for $i=\,$I and IV
the combination $\cT^{[i]}_\alpha+\frac{\I}{2}\,\bG_{\rm IIB}(\xi)$
is given by the same expression
\eqref{diffTG}. Thus, it is also holomorphic and removable by
a gauge transformation in the patch $\cU_{\bm}$, so that in both cases one recovers
the transition functions of the Type IIB formulation.

Finally, it remains to consider the transition functions between the north pole and different quadrants.
They are given by
\be
H^{[+i]_B}=\frac{\I}{2}\, F+\frac{\I}{2}\,\cG\ui{i}-\cT^{[i]}_\alpha\, ,
\label{transB+i}
\ee
where $\cG\ui{i}$ denotes the corresponding holomorphic branch of the function \eqref{funGg}
in $\cU_i$. The last two terms coincide with $\frac{\I}{2}\,\cG-\cT_\alpha$.
This combination decomposes into two parts. One part is the ``constant" term
$\frac{\I \ve \tve}{192} \, \chi_X$ with discontinuities along real and imaginary axes,
while the rest is
holomorphic\footnote{In fact, the situation is complicated by the cut along the real axis
originating from dependence of $\Delta_\gamma$ on $\tilde\ve$ in \eqref{eq:master}. However,
the entire real axis (except 0) is supposed to be covered by $\cU_{\bm}$ and $\cU_{\bp}$ so that
formally our function is indeed regular in the domain under consideration. Another way to see
this is to use the representation \eqref{eq:A21} where, instead of the cut, one has
infinitely many poles which by assumption belong to $\cU_{\bm}$ and $\cU_{\bp}$.
All these complications are in fact a consequence of that the covering used to define
the Type IIB twistor space is not regular (see footnote \ref{foot_cover}).}
in $\cU_+\setminus (\cU_{\bm}\cup\cU_{\bp})$ and therefore can be removed
by an appropriate gauge transformation.
As a result, up to the above ``constant" term, the transition functions between
the north pole and all quadrants are given by the holomorphic prepotential only.
A similar gauge transformation can be performed
in the patch $\cU_-$ around the south pole where one finds the same result.

The remaining constant piece then has precisely the correct form to
cancel the anomalous dimension which is taken to be vanishing on the Type IIB side.
To see this, let us note that the presence of singular
terms in the twistor lines due to anomalous dimensions allows for similar singularities
in the gauge transformations. In particular, the gauge transformation generated by
$\Ti{i}=-\frac{\I \ve \tve}{192} \, \chi_X$ for {\it all} patches simply rotates
the cuts of the logarithm from the imaginary to the real axis. In our case,
such a gauge transformation has been already performed in the four quadrants
and, once it is done in $\cU_\pm$, it precisely cancels the remaining constant terms in $H^{[\pm i]_B}$.
However, this argument requires this gauge transformation to be performed also in $\cU_{\bmp}$,
which would introduce additional terms in $H^{[i \bmp]_B}$.
Setting these terms to zero ``by hand'' is equivalent to taking
the anomalous dimension to be vanishing. Thus we
arrive at the same set of anomalous dimensions
and transition functions \eqref{symp-inst}
which describes the twistor space of Type IIB HM moduli space.

At the end,
we recapitulate the gauge transformations which map the initial Type IIA twistor space
into the Type IIB one. They read
\be
T^{[i]} =   \cT^{[i]}_\alpha \, ,
\quad
T^{[\pm]} = \cT_\alpha  \mp \frac{\I}{2}\, \cG \, ,
\quad
T^{[\bp]} = \cT_\alpha +\frac{\I}{2}\, G_{\rm IIB}\, ,
\quad
T^{[\bm]} = \cT_\alpha + \frac{\I}{2}\, \bG_{\rm IIB}  \, .
\label{allagauge}
\ee
This completes the proof and establishes mirror symmetry between
the Type IIA and Type IIB descriptions at the level of the twistor space.

\section{Poisson resummation of the twistor lines}
\label{sec_Pois}

In this appendix we collect the technical details underlying the Poisson resummation of the
twistor lines \eqref{xiqlineB} with respect to the quantum number $q_0$, thereby
providing the derivation of eq.\ \eqref{IIAtw}. Inspecting \eqref{xiqlineB}, we notice that
all instanton corrections are essentially encoded in the functions $\Ikl^{(\nu)}(\varpi)$.
Since the instanton numbers $\hnkl$ are independent of $q_0$, one can simplify the calculation
by first considering the Poisson resummation of
\bse
\label{basicresum}
\bea
\label{defcI}
\cI^{(\nu)}_{\hat\gamma}(\varpi) \equiv \sum_{q_0\in\IZ} \Ikl^{(\nu)}(\varpi)
\quad {\rm \ with\ } \quad \hat\gamma\ne 0 \, , \\
\label{defcII}
\cI^{(2)}_0(\varpi) \equiv
\sum_{q_0\ne 0} \cI_{\gamma}^{(2)}(\varpi) \quad {\rm \ with\ } \quad \hat\gamma=0,
\eea
\ese
where we have extracted the sum over the charge $q_0$ from the lattice
sum over $\gamma=\{q_0,\hat\gamma\}$. Given these ``basic resummations''
the other terms entering the twistor lines can be obtained by differentiating
with respect to appropriate fields.
Thus, we first consider the resummation
of \eqref{basicresum} before applying the result to the twistor lines
in Subsection \ref{App:A2}.

\subsection{Resuming $\Ikl^{(\nu)}(\varpi)$}

In general Poisson resummation relies on the formula
\be
\sum\limits_{n\in \IZ} f(x+n)= \sum\limits_{n\in \IZ}\tilde f(2\pi n)e^{2\pi i n x},
\label{Pois}
\ee
where
\be
f(x)=\frac{1}{2\pi}\int^{\infty}_{-\infty} dw \, \tilde f(w) e^{iwx},
\qquad
\tilde f(w)=\int^{\infty}_{-\infty} dx \, f(x) e^{-iwx} \, ,
\ee
are related by Fourier transform.

In order to apply this formula to \eqref{defcI}, we introduce the following notations
\be
x=q_a b^a,
\qquad
y=q_a t^a,
\qquad
\Thhg= q_a (\zeta^a-b^a\zeta^0) \, .
\label{identPois}
\ee
Then the function $f(x + q_0)$ entering the l.h.s. of \eqref{Pois} is found as
\be
f(x)=\sum_{m=1}^{\infty} \sum_{s=\pm 1}
\frac{e^{-2\pi \I s m(\Thhg+x\zeta_0 ) }}{(s m)^\nu}
\int_{0}^{\infty}\frac{\d t}{t}\, \frac{t-\sign(x) s\I \varpi}{t+\sign(x) s\I\varpi}\,
e^{-2\pi m\sign(x)\cR \( \frac{x+\I y}{t} +t (x-\I y)\)}.
\label{funP}
\ee
Its Fourier transform is obtained by using $\sign(x)$ to split the $x$-integration
into two integrals along the half-axes. The resulting integrands are
of the type $e^{-Ax}, {\rm Re}A > 0$. Their evaluation yields
\be
\tilde f(w)=\frac{1}{2\pi}\sum_{m=1}^{\infty} \sum_{s,s'=\pm 1}
\frac{e^{-2\pi \I s m \Thhg }}{(s m)^\nu}
\int_{0}^{\infty}\frac{\d t}{t}\, \frac{t-s s'\I \varpi}{t+s s'\I\varpi}\,
\frac{e^{-2\pi \I m s' \cR y \( t^{-1} -t\)}}{m \(\cR \( t^{-1} +t\)+\I s s' \zeta^0\) +\frac{\I s'w}{2\pi}}.
\ee
The sign variables $s$ and $s s'$ can then be used to
extend the sum over $m$ to negative values and the $t$-integral along the
whole real axis, respectively.
This should be contrasted with eq.\ \eqref{funP}
where the extension of the $t$-integral to the real axis cannot be performed.
Substituting this result into the r.h.s.\ of \eqref{Pois} then yields
\be
\cI^{(\nu)}_{\hat\gamma} = \frac{1}{2\pi}\sum_{n \in \IZ \atop m\ne 0 }
\frac{e^{2\pi \I (n x - m \Thhg) }}{m^{\nu-1}|m|}
\int_{-\infty}^{\infty}\frac{\d t}{t}\, \frac{t-\I \varpi}{t+\I\varpi}\,
\frac{e^{-2\pi \I m  \cR y \( t^{-1} -t\)}}{m \(\cR \( t^{-1} +t\)+\I \zeta^0\) + \I n}.
\ee
The $t$-integral can now be evaluated explicitly using the method of
residues by closing the integration contour at infinity in the half-plane
where the integrand is exponentially suppressed. The integrand has
three poles located at $t=-\I\varpi$ and $t=-\I\varpi_\pm^{m,n}$
with $\varpi_\pm^{m,n}$ defined in \eqref{poles}. Their contribution depends on the signs
\be
\label{defeps}
\eps=\sign y \, , \qquad \ve=-\sign(\Re \varpi) \, .
\ee
For $\eps = \pm 1$ the contour includes the poles at $\varpi^{m,n}_\pm$, respectively.
The pole at $t=-\I\varpi$ contributes for $my\Re \varpi <0$ only,
and yields different contributions on the four quadrants in the $(y,\Re\varpi)$-plane.
Introducing sign-functions in the appropriate places the resulting expressions can be combined to
\be
\begin{split}
\cI^{(\nu)}_{\hat\gamma}=&
\sum_{ n \in \IZ \atop m \not = 0  }\frac{e^{2\pi i( n x-m \Thhg )-2\pi |y| |m\tau +n|}}{m^\nu|m\tau+n|}\,
\frac{\varpi_\eps^{m,n}+\varpi}{\varpi_\eps^{m,n}-\varpi}
+2\eps \sum_{ n \in \IZ \atop \eps\ve m>0  }
\frac{e^{2\pi i( n x- m \Thhg )}}{m^\nu}\,
\frac{e^{2\pi  m y \cR \(\varpi^{-1}+\varpi\)}}{m \xi^0 +n}.
\end{split}
\label{cIhat}
\ee
This result completes the resummation of $\cI^{(\nu)}_{\hat\gamma}$.

We now turn to the second ``basic resummation'' \eqref{defcII}.
Substituting \eqref{IKkl}, it explicitly reads
\be
\cI^{(2)}_0(\varpi) =
\sum_{q_0\ne 0}\sum_{m=1}^{\infty} \sum_{s=\pm 1} \frac{e^{-2\pi \I s m q_0 \zeta^0}}{m^2}
\int_{0}^{\infty}\frac{\d t}{t}\, \frac{t-\sign(q_0) s\I \varpi}{t+\sign(q_0) s\I\varpi}\,
e^{-2\pi m |q_0| \cR\( t^{-1}  +t \)}.
\label{exprcIzero}
\ee
The Poisson resummation formula then requires to include the $q_0=0$-term in the sum
\be
\cI^{(2)}_0(\varpi) =
\sum_{q_0\in \IZ}f(q_0)-f(0) \, .
\label{exprcIz}
\ee
For $q_0 = 0$, the $t$-integral arising from \eqref{exprcIzero} can be evaluated analytically.
In fact, at every boundary it diverges logarithmically, but these divergences can be canceled
by first combining contributions of $t$ and $1/t$ since this transformation leaves
the exponential invariant. Furthermore,
the factors $\sign(q_0)$ can be removed by redefining $s \rightarrow \sign(q_0) s$, which leads
to $|q_0|$ in the first exponential. The functions entering into \eqref{exprcIz} then read
\be\label{Ffunz}
\begin{split}
f(q_0)= & \, \sum_{m=1}^{\infty} \sum_{s=\pm 1} \frac{e^{-2\pi \I s m |q_0| \zeta^0}}{m^2}
\int_{0}^{\infty}\frac{\d t}{t}\,\frac{t- s\I \varpi}{t+ s\I\varpi}\,
e^{-2\pi m |q_0| \cR\( t^{-1}  +t \)} \, , \\
f(0) = & \, - \frac{2 \pi^2}{3} \, \log \varpi \, .
\end{split}
\ee
Note that $f(q_0)$ is obtained from \eqref{funP} by setting $q_a=0$. Its Poisson resummation
can be obtained by following similar steps as for $\cI^{(\nu)}_{\hat\gamma}$.
The result turns out to be the same as \eqref{cIhat} evaluated for $\nu=2$, $q_a = 0$ and
$\eps=1$, which can be verified by an explicit evaluation of the integrals.
Substituting it into \eqref{exprcIz}
then yields
\be
\begin{split}
\cI^{(2)}_0(\varpi) & =
\sum_{n\in \IZ \atop m\ne 0}
\frac{1}{m^2 |m \tau + n|} \, \frac{\varpi^{m,n}_+ + \varpi}{\varpi^{m,n}_+ - \varpi}
+  \sum_{n\in \IZ \atop m> 0} \frac{2 \ve}{m^2 (m \xi^0+n)}
+\frac{2\pi^2}{3}\, \log\varpi .
\end{split}
\label{rescIz}
\ee
Notice that the second sum over $n$ converges because the combination
of $n$th and $-n$th terms scales like $n^{-2}$.
This result concludes the Poisson resummation of \eqref{basicresum} and
we will now proceed with its application to the Type IIA twistor lines.

\subsection{Resuming the Type IIA twistor lines}
\label{App:A2}

Under the decomposition $\gamma = \{q_0, \hat{\gamma}\}$ the instanton corrections
to the twistor lines \eqref{xiqlineB} give rise to four types of terms:
$q_a \cI^{(1)}_{\hat\gamma}$ linear in charges, $\cI^{(2)}_{\hat\gamma}$ without charges,
$q_0 \cI^{(1)}_{\gamma}$ linear in $q_0$, and terms with $\cI^{(2)}_{0}$.
By virtue of the relation $\cK_\gamma = \frac{\I}{4} \cI_\gamma^{(1)}(0)$, the contributions
of the form $q_a \cK_\gamma$ and $q_0 \cK_\gamma$ result as a special case.

Taking into account the sum over charges and that
$\hnkl=n_{\hat\gamma}=n_{-\hat\gamma}$ for $\hat\gamma\ne 0$,
the expressions for the first two types of terms can be summarized as
\be
\begin{split}
\sum_{\hat\gamma\ne 0} n_{\hat\gamma} q_a^{2-\nu} \cI^{(\nu)}_{\hat\gamma}= &\,
2 \sum_{\hat\gamma_+} n_{\hat\gamma} q_a^{2-\nu} \bigg[ \sum\limits_{n\in \IZ \atop m\ne 0}
\frac{\e^{2\pi i( n x-m \Thhg )}}{m^\nu}\,
\frac{\e^{-2\pi y |m\tau +n|}}{|m\tau+n|}\,
\frac{\varpi_+^{m,n}+\varpi}{\varpi_+^{m,n}-\varpi}
 \\
&
\qquad \qquad + 2 \varepsilon^{\nu+1} \sum\limits_{n\in \IZ \atop m > 0}
\frac{\e^{2\pi \varepsilon ( m y \cR \(\varpi^{-1}+\varpi\)
+ \I ( n x- m \Thhg ))}}{m^\nu \, \( m \xi^0 +n \)}\,
\bigg].
\end{split}
\label{cIhatsum}
\ee
Here $\hat\gamma_+$ is the set of charges satisfying $y >0$ and
already appeared in the description of Section \ref{sec_instA}.
The terms linear in $q_0$ are related to this result by
\be
\sum_{q_0\in\IZ} q_0 \Ikl^{(\nu)}(\varpi)=-\frac{1}{2\pi \I}\, \p_{\zeta^0}\cI^{(\nu+1)}_{\hat\gamma} \, ,
\label{dercI}
\ee
which can be established based on the definition \eqref{IKkl}
where the r.h.s.\ is expressed in terms of Type IIA variables.
Finally, either setting $\varpi=0$ in the above equations or redoing calculations
from the very beginning, one can show that
\be\label{Ksum}
\begin{split}
&\sum\limits_{\hat{\gamma} \ne 0} n_{\hat{\gamma}} q_a \sum_{q_0\in \IZ}\Kkl
=  \frac{\I}{2} \sum_{\hat\gamma_+} n_{\hat\gamma} q_a \sum_{n\in \IZ \atop m\ne 0}
\frac{e^{2\pi i( n x-m \Thhg )-2\pi |y| |m\tau +n|}}{m|m\tau+n|},
\\
&\sum_{\gamma}\hnkl q_0\Kkl = -\frac{1}{4\pi}\sum_{n\in \IZ \atop m\ne 0}
\left[\sum_{\hat\gamma_+}n_{\hat\gamma}\p_{\zeta^0}
\frac{e^{2\pi i( n x-m \Thhg )-2\pi y |m\tau +n|}}{m^2|m\tau+n|}
+\frac{\chi_Y}{2}\, \frac{m\tau_1+n}{m|m\tau+n|^3} \right].
\end{split}
\ee

With these relations, we now have all the ingredients to compute
the resummed Type IIA twistor lines \eqref{xiqlineB} and compare them to
their Type IIB counterparts \eqref{IIBtwist}. Since one starts from the Type IIA side,
we work in the patch $\cU_{0_A}$.
The decomposition of the sum over charges in $\txi_\Lambda^{[0_A]}$ then yields
\be\label{U0A}
\begin{split}
\txii{0_A}_a = &
\frac{\I}{2}\(\tzeta_a
+\cR \left( \varpi^{-1} F_a - \varpi \, \bF_a \right)\)
+\frac{\I}{32\pi^2}
\sum_{\hat\gamma\ne 0} n_{\hat\gamma} q_a \cI^{(1)}_{\hat\gamma} \, ,
\\
\txii{0_A}_0 = &
\frac{\I}{2}\(\tzeta_0
+\cR \left( \varpi^{-1} F_0 - \varpi \, \bF_0 \right)\)
-\frac{1}{64\pi^3}\,\p_{\zeta^0}\(\sum_{\hat\gamma\ne 0} n_{\hat\gamma}
\cI^{(2)}_{\hat\gamma}-\chi_Y\cI^{(2)}_{0}\) \, .
\end{split}
\ee
In order to facilitate the comparison between the last twistor lines, it turns out to be convenient
to trade $\aA^{[0_A]}$ for the combination $\aB^{[0_A]}$ given in \eqref{defha}. For the latter
quantity the decomposition gives
\be
\begin{split}
\aB^{[0_A]}
=& -\frac{\I}{4}\Big[\sigma + \cR ( \varpi^{-1} W - \varpi \bar{W})
\\
& \qquad \, \,- \(\zeta^\Lambda + \cR( \varpi^{-1}z^\Lambda  - \varpi \zb^\Lambda )\)
\big(\tilde{\zeta}_\Lambda + \cR(\varpi^{-1} F_\Lambda  - \varpi \bar{F}_\Lambda )\big) \Big] \\
&  -\frac{\chi_Y}{96\pi}\,\log \varpi
- \frac{1}{64 \pi^3} \Big[
\sum\limits_{\hat{\gamma} \ne 0}  n_{\hat{\gamma}} \cI^{(2)}_{\hat{\gamma}} - \chi_Y \cI^{(2)}_{0}
\Big] \\
& + \frac{\cR}{8 \pi^2} \Big[ \sum\limits_{\hat{\gamma} \ne 0}  n_{\hat{\gamma}}  q_a (\varpi^{-1} z^a
+ \varpi \zb^a) \sum\limits_{q_0 \in \IZ} \cK_\gamma +
(\varpi^{-1} + \varpi ) \sum\limits_{\gamma} n_{\gamma} q_0 \cK_\gamma \Big] \, .
\end{split}
\ee
Substituting the relations \eqref{rescIz} - \eqref{Ksum}, a straightforward though somewhat tedious
computation allows to establish the following relation between the Type IIA and Type IIB twistor lines
\be\label{reltwi}
\txii{0_A}_\Lambda = \txii{0_B}_\Lambda + \tilde\cT_\Lambda \, ,
\qquad \aB^{[0_A]} = \aB^{[0_B]} + \cT_\alpha\, ,
\ee
where we introduced
\be\label{eq:A21}
\begin{split}
\tilde\cT_a = & \, \frac{\I}{8 \pi^2} \sum\limits_{q_a \ge 0} n^{(0)}_{q_a} q_a
\sum\limits_{n \in \IZ \atop m > 0}
\frac{\e^{-2 \pi \I \ve q_a \( m \xi^a - b^a (m \xi^0 + n)\)} }{m \( m \xi^0 + n \)} \, ,
\\
\tilde\cT_0 = & \, \frac{1}{16 \pi^3} \sum\limits_{q_a \ge 0} n^{(0)}_{q_a} \sum\limits_{n \in \IZ \atop m > 0}
\( \ve - 2 \pi \I q_a b^a \( m \xi^0 + n \) \)
\frac{\e^{-2 \pi \I \ve q_a \( m \xi^a - b^a (m \xi^0 + n)\)} }{m \( m \xi^0 + n \)^2} \, ,
\\
\cT_\alpha = & \, - \frac{\ve}{16 \pi^3} \sum\limits_{q_a \ge 0} n^{(0)}_{q_a}
\sum\limits_{n \in \IZ \atop m > 0}
\frac{\e^{-2 \pi \I \ve q_a \( m \xi^a - b^a (m \xi^0 + n)\)} }{m^2 \( m \xi^0 + n \)} \, .
 \end{split}
\ee
In order to verify these identities, it is useful to note that the terms proportional
to $\chi_Y$ provide the $q_a = 0$ part of the sums over charges
once \eqref{instnr} is applied. Furthermore, we have
\be
\begin{split}
2 \pi y |m \tau + n|-2 \pi \I (n x - m \Theta) = & \,  S_{m,n,q_a} \, ,
\\
-m y \cR ({\bf z}^{-1} + {\bf z}) - \I \(n x - m \Theta\) = & \, \I q_a \( m \xi^a -  b^a (m \xi^0 + n ) \) \, .
\end{split}
\ee
Tracing back the origin of the terms appearing in \eqref{reltwi}, it is worthwhile noting
that the twistor lines $\txii{0_B}_\Lambda, \aB^{[0_B]}$ are generated by the poles
$t = -\I \varpi^{m,n}_\pm$ while the $\cT$-terms originate from the pole at $t = -\I \varpi$.

In fact, it can be demonstrated that away from the real and imaginary axes,
the additional contributions $\tilde{\cT}_\Lambda, \cT_\alpha$ are holomorphic functions
of $\xi^\Lambda$. This is a crucial prerequisite for removing these terms by
an appropriate gauge transformation.
To show this, one should ``undo'' the Poisson resummation on $n$ for these extra contributions.
In this course, we need the following relations in the sector $q_a \ne 0$
\bea\label{res:int1}
&&\sum_{n\in \IZ}\frac{\e^{  2\pi\I \ve q_a   b^a (m\xi^0+n) }}{  m \xi^0 +n} =
2 \pi \I  \ve  \sum\limits_{q_0 \in \IZ} \Delta_{\gamma}(\varpi) \e^{-2 \pi \I \ve m q_0 \xi^0}\, ,
\\
&& \sum_{n\in \IZ}\(1-   2\pi \I \ve q_a b^a \( m \xi^0 +n\)\)
\frac{\e^{  2\pi\I \ve q_a  b^a(m\xi^0+n)}}{\( m \xi^0 +n\)^2} =
-4\pi^2\sum_{q_0\in \IZ}q_0  \Delta_{\gamma}(\varpi)\,\e^{-2\pi \I \ve m q_0 \xi^0}\, ,
\nonumber
\eea
where we introduced the step-function
\be
\Delta_{\gamma}(\varpi)=\left\{
\begin{array}{llll}
0  \quad & {\rm if} \quad \epskl \Re\varpi\, \Im \varpi >0 &
\quad \Leftrightarrow \quad & \eps_\gamma \ve \tve= -1
\\
\epskl  \quad & {\rm if} \quad \epskl \Re\varpi\, \Im \varpi <0 &
\quad \Leftrightarrow \quad & \eps_\gamma \ve \tve= 1
\end{array}
\right.
\ee
and $\tve=\sign(\Im\varpi)$.
These relations can be obtained in the usual way by performing Poisson resummation
which boils down to the evaluation of continuous Fourier transform w.r.t. the variable $n$.
The latter is found by closing the integration contour in the upper (lower) half-plane
where the integrand is exponentially suppressed. Taking into account that $m>0$,
the analysis of the pole-structure reveals that the integral is non-zero
for $\eps_\gamma \ve \tve= 1$ and vanishes otherwise,
which leads to the appearance of the step-function.
In the sector $q_a = 0$, a similar resummation gives
\be\label{id3}
\begin{split}
\sum_{n\in \IZ}\frac{1}{m\xi^0+n}=& \sum_{n\in \IZ}\frac{m\xi^0}{(m\xi^0)^2-n^2}
=2\pi \I \tve \left( \sum_{q_0=1}^\infty \e^{- 2\pi \I \tve m q_0\xi^0} + 1/2 \right) ,
\\
\sum_{n\in \IZ}\frac{1}{(m\xi^0+n)^2}=& \sum_{n\in \IZ}\frac{(m\xi^0)^2+n^2}{\((m\xi^0)^2-n^2\)^2}
=-4\pi^2 \sum_{q_0=1}^\infty q_0\,\e^{- 2\pi \I \tve m q_0\xi^0} \, .
\end{split}
\ee

Substituting the identities \eqref{res:int1}, \eqref{id3} into eqs.\ \eqref{eq:A21}
and carrying out the summation over $m$ explicitly then leads to our final result
\be
\begin{split}
\tilde{\cT}_a = & \, \frac{\ve}{4\pi} \sum\limits_{q_0 \in \IZ,\,q_a\ge 0  } n_{q_a}^{(0)}
q_a \Delta_{\gamma}(\varpi)  \log\( 1 - \e^{-2 \pi \I \ve q_\Lambda \xi^\Lambda} \) \, ,
\\
\tilde{\cT}_0 = & \, \frac{\ve}{4\pi} \sum\limits_{q_0 \in \IZ,\,q_a\ge 0  } n_{q_a}^{(0)}
q_0 \Delta_{\gamma}(\varpi)  \log\( 1 - \e^{-2 \pi \I \ve q_\Lambda \xi^\Lambda} \)
\, ,
\\
\cT_\alpha = &\,
- \frac{\I}{8 \pi^2} \sum\limits_{q_0 \in \IZ,\,q_a\ge 0  \atop \gamma\ne 0}
n_{q_a}^{(0)}  \Delta_{\gamma}(\varpi) \Li_2\(\e^{-2 \pi \I \ve q_\Lambda \xi^\Lambda}\)
- \frac{\I \ve \tve}{192} \, \chi_X \, .
\end{split}
\label{eq:master}
\ee
In the regions where the step-function $\Delta_\gamma$ is constant,
the $\cT$-contributions indeed depend only on the twistor lines $\xi^\Lambda$ in a holomorphic way.
Moreover, it is straightforward to see that,
away from the discontinuities induced by $\ve$ and $\tve$,
$\tilde{\cT}_\Lambda = \p_{\xi^\Lambda} \cT_\alpha$,
which is a necessary requirement for the extra terms to constitute a gauge transformation.

\section{SL(2,$\IZ)$-transformation of the Type IIB twistor lines}
\label{ap_transform}

The key ingredient in the construction of the non-perturbative mirror map
\eqref{instmap} is the fact that the SL(2, $\IZ$)-transformation
of the Type IIB twistor lines \eqref{newtxinew} -
\eqref{twhatalpha} is given by the classical
transformation law \eqref{SL2Zxi}.
In this appendix we provide the details underlying
the derivation of this result.

First, it is useful to decompose the transformation of $\tau = \tau_1 + \I \tau_2$,
eq.\ \eqref{SL2Z}, into its real and imaginary part
\be
\tau_2\mapsto \frac{\tau_2}{|c\tau+d|^2} \, ,
\qquad
\tau_1\mapsto \frac{ac|\tau|^2+bd+(ad+bc)\tau_1}{|c\tau+d|^2} \, .
\ee
Moreover, using the notation $\varpi_\pm^{c,d}$ introduced in \eqref{poles},
the transformation of $\varpi$ \eqref{transz} can be expressed as
\be
\varpi  \mapsto  \frac{1+\varpi_-^{c,d}\varpi}{\varpi_-^{c,d}-\varpi}=
-\frac{\varpi_+^{c,d}-\varpi}{1+\varpi_+^{c,d}\varpi} \, ,
\ee
which implies
\be
\varpi^{-1}+\varpi \mapsto  \frac{|c\tau+d|}{c\xi^0+d}\(\varpi^{-1}+\varpi\) \, ,
\qquad
\varpi^{-1}-\varpi \mapsto  \frac{(c\tau_1+d)\(\varpi^{-1}-\varpi\)-2c\tau_2}{c\xi^0+d}.
\ee

With regards to the transformation of the instanton-sums appearing
in the twistor lines, it is useful to introduce
\be
\( m'\atop n'\) =
\(
\begin{array}{cc}
a & c
\\
b & d
\end{array}
\)
\( m \atop n \) \, .
\ee
Utilizing this notation, one then proves the transformation
properties
\be\label{gen:traf}
S_{m,n,q_a}  \mapsto   S_{m',n',q_a} \, ,
\qquad
|m\tau+n| \mapsto  \frac{|m'\tau+n'|}{|c\tau+d|} \, ,
\ee
which enter into all twistor lines.
With these preliminary results, we are now in the position to discuss
the transformation of $\txi_a, \txi_0$, and $\aB$ in turn.

We start by considering $\txi_a$. Noting that the
second term appearing in \eqref{SL2Zxi} is already generated by the classical
piece in the first line of eq.\ \eqref{newtxinew}, it follows that the instanton
contribution in the second line must be modular invariant.
That this is indeed the case readily follows from \eqref{gen:traf} together
with the invariance of
\be
\frac{1+\varpi_+^{m,n}\varpi}{\varpi-\varpi_+^{m,n}}
\mapsto
\frac{1+\varpi_+^{m',n'}\varpi}{\varpi-\varpi_+^{m',n'}} \, .
\ee
Concerning the transformation of $\txi_0$, we first verify that the classical pieces
given by the first line of \eqref{newtxizeronew} satisfies the classical \eqref{SL2Zxi}.
Hence the transformation of the instanton contributions given by the second and third line
has to reproduce the instanton pieces originating from $(c \xi^0 + d) \txi_0 - c (\aB - \xi^a \txi_a)$.
This can be verified by applying the identities
\be
\begin{split}
\frac{m \tau_1 + n}{|m \tau + n|^2} & \mapsto
\frac{d (m' \tau_1 + n') + c (n' \tau_1 + m' |\tau|^2)}{|m'\tau+n'|^2}\, ,
\qquad
m\xi^0+n   \mapsto  \frac{m'\xi^0+n'}{c\xi^0+d} \, ,
\\
& \frac{1 - \varpi_+^{m,n} \varpi}{\varpi - \varpi_+^{m,n}} \, t^a
 \mapsto
(c \tau_1 + d) \, \frac{1 - \varpi_+^{m',n'} \varpi}{\varpi - \varpi_+^{m',n'}} \, t^a
- c  \, \frac{\varpi + \varpi_+^{m',n'}}{\varpi - \varpi_+^{m',n'}} \, \tau_2 \, t^a \, .
\end{split}
\label{1.2ex}
\ee
Finally, following the same strategy as for $\txi_0$, on verifies the transformation
law of $\aB$ by first establishing the intermediate result
\be
(m\tau_1+n)\(\varpi^{-1}-\varpi\)-2m\tau_2  \mapsto
\frac{(m'\tau_1+n')\(\varpi^{-1}-\varpi\)-2m'\tau_2}{c\xi^0+d} \, .
\ee
This completes the proof that the classical SL(2, $\IZ$) transformations \eqref{SL2Zxi}
do not receive quantum corrections from D($-1)$ and D1-brane instantons.


\providecommand{\href}[2]{#2}\begingroup\raggedright\endgroup

\end{document}